\documentclass{aastex63} 
\usepackage{amsmath}
\usepackage{amssymb}
\usepackage[normalem]{ulem}
\usepackage{appendix}
\usepackage{graphicx}

\begin{document}
\shorttitle{Cepheus X-4 observations with AstroSat}
\shortauthors{Mukerjee et al.}
\title{Studies of Cepheus X-4 during 2018 outburst observed with AstroSat}
\author[0000-0003-4437-8796]{Kallol Mukerjee}
\affiliation{Tata Institute of Fundamental Research, Homi Bhabha Road, Mumbai 400005, India}
\email{kmukerji@tifr.res.in}
\author[0000-0001-7549-9684]{H. M. Antia}
\affiliation{Tata Institute of Fundamental Research, Homi Bhabha Road, Mumbai 400005, India}
\affiliation{UM--DAE Centre for Excellence in Basic Sciences, University of Mumbai, Mumbai 400098, India}
\email{antia@tifr.res.in}

\begin{abstract}

We present timing and spectral results of 2018  outburst of  Cepheus X-4, observed twice by AstroSat  
at luminosity of $2.04 \times 10^{37}$ erg s$^{-1}$  and  $1.02 \times 10^{37}$ 
erg s$^{-1}$ respectively. The light curves showed strong pulsation and co-related X-ray intensity variation 
in SXT (0.5--8.0 keV) and  LAXPC (3--60 keV) energy bands. Spin-period and spin-down rate  of the pulsar 
were determined from two  observations  as  
$65.35080\pm0.00014$ s , $(-2.10\pm0.8)\times10^{-12}$ Hz s$^{-1}$ at  an epoch MJD 58301.61850 and
$65.35290\pm0.00017$ s, $(-1.6\pm0.8)\times10^{-12}$ Hz s$^{-1}$ for an epoch
MJD 58307.40211.  Pulse-shape studies with AstroSat showed energy and intensity dependent variations. The pulsar 
showed an  overall  continuous spin-down, over 30 years  at an average-rate  of $(-2.455\pm0.004)\times10^{-14}$ 
Hz s$^{-1}$, attributed  to propeller-effect in the subsonic-regime of the pulsar,  
in addition to variations during  its  outburst activities.
Spectra  between 0.7--55 keV  energy band were well fitted by  two continuum  
models,  an absorbed  compTT-model and an absorbed power-law with a Fermi-Dirac cutoff (FD-cutoff) model with a 
black-body. These were  combined with  an iron-emission line and a cyclotron absorption line.  The prominent 
cyclotron resonance  scattering features with a peak absorption energy of $30.48^{+0.33}_{-0.34}$ keV and 
$30.68^{+0.45}_{-0.44}$ keV  for FD-cutoff-model and $30.46^{+0.32}_{-0.28}$ keV and $30.30^{+0.36}_{-0.34}$ keV 
for compTT-model were detected during two AstroSat observations.  These when compared with earlier results, 
showed  long term stability of its  average value of $30.23 \pm 0.22$ keV. The pulsar showed  pulse-phase  as 
well as luminosity dependent variations in  cyclotron-line energy and width  and in plasma optical-depth 
of its spectral continuum.

\end{abstract}

\section{Introduction}

X-ray pulsars offer vital  information through studies of their pulse characteristics about 
their nature  and   geometry of the binary system.  The shape of the pulse profile   offers  
insight about mode of accretion inflows, its luminosity, geometry of accretion column and 
its  magnetic field configuration \citep{parmar89a}.  Therefore, such studies offer understanding of  
its pulsar system and the process of mass accretion in the presence of strong magnetic field of the neutron star. 
Detailed studies on  pulse characteristics  were conducted during outburst activities  of some of the 
pulsars earlier  such as EXO 2030+375 \citep{parmar89a}, Cepheus X-4 \citep{koyama91,mihara91,mukerjee00},
XTE J1946+274 \citep{paul01} etc., which offered valuable information and 
understanding of physical processes responsible for the  observed properties of these pulsars.
Spectroscopic studies,  on the other hand, reveal  environment surrounding the pulsar and the underlying 
mechanism  responsible for energy generation  in these systems.  Detailed studies on cyclotron absorption 
features, if present in the pulsar spectrum, not only enable us to  determine its surface magnetic field,
but also offers an insight into the line producing region, its structure and geometry of the accretion column 
\citep{staubert19}. Therefore, detailed studies of cyclotron absorption features are being pursued  by 
researchers since its discovery \citep{truemper78} as it is an important diagnostic probe for pulsars. 
Cyclotron absorption features  were detected in the spectrum of many Be-binaries covering  a wide range 
of energies, starting from a lower energy of $\sim10$ keV \citep{decesar13, jun12} to  a higher energy 
at $\sim100$ keV \citep{barbara01}.  Detailed studies of cyclotron line sources and their properties were 
reported by \citet{staubert19} and \citet{maitra17}.  The Be-binaries,  particularly because of their 
transient nature,  offer us opportunities of detailed studies of some of these interesting properties 
with change in source luminosity and in time. The studies on cyclotron-line energy with respect to 
its pulse-phase, source luminosity and with time, showed a wide variations for some sources, such as 
Vela X-1, Cen X-3 and Her X-1 \citep{staubert19}.  These interesting properties help us to systematically 
investigate and understand their underlying physical properties. 

A Be-binary  pulsar system is  a class of high mass X-ray binary,  
where a neutron star  has a  Be-star as its companion,  shows significant variations 
of the source luminosity and pulse period during its outburst activity. 
The stellar wind from rapidly rotating Be-stars  are highly asymmetric 
and frequently form equatorial accretion disc around the Be-star. Such systems 
typically have a large orbital eccentricity resulting from relatively recent supernova explosion. 
The presence of these two features, equatorial disc and large eccentricity, enables the matter to 
accrete on to the neutron star directly from the equatorial Be-star disc during its
periastron passage. This  results in a short duration  outburst  due to  enhanced accretion on to the
neutron star and cause increase in the observed X-ray flux by several orders of 
magnitude, typically up to $\approx 10^{37}$~erg~s$^{-1}$. Such outburst events occur in a particular
phase of the orbit and are  identified as a Type-I outburst. 
In addition to such events, there are Type-II outbursts, which are due to a sudden 
increase in amount of matter in the circumstellar disc surrounding the Be-star, which can occur 
at any orbital phase \citep{okazaki01, okazaki02, okazaki16}. 
The duration of Type-II outbursts varies from weeks to months, during 
which the source X-ray luminosity can reach the Eddington limit, $\approx 10^{38}$ erg~s$^{-1}$ Reig (2011).
Therefore, Be-binary systems in particular, offer the best  opportunities for studies of variation of
pulse characteristics and spectral behavior of sources with change in source luminosity during its 
outburst activity.  Hence, many Be-binaries  were observed over the period during outbursts and  
offered interesting results for understanding accretion theory  \citep{bildsten97, reig11}.
The Cepheus X-4 is one such interesting Be-binary pulsar which showed
rare and long outburst of about 40 days in a span of about 4--5 years and  was observed and  studied
by many observatories from time to time during its outburst activities.

Cepheus X-4  was discovered by X-ray telescope of OSO-7 as a 
transient source  during June--July, 1972 \citep{ulmer73}.  Ginga 
observed the source during March 1988 outburst and detected spin period of 
66.25 s of its neutron star for the first time \citep{koyama91}.
Spectroscopic studies from the same Ginga observations lead to detection 
of cyclotron resonance scattering feature corresponding to a centroid
energy at $30.5\pm0.4$ keV \citep{mihara91}.
The  ROSAT observations during  June 1993 outburst refined the source 
coordinates and also determined its pulsar spin period \citep{schulz95}. 
The  observations by BATSE during June--July 1993 and  during subsequent
outburst of June--July 1997 which was also followed by RXTE, determined 
pulse characteristics of Cepheus X-4. Additionally, a  possible range of its orbital period 
from 23--147.3 days was suggested  using RXTE data \citep{wilson99}.  
From  observed  characteristic features and outburst activities, it was  predicted that  
Cepheus X-4 could possibly have a massive early type Be-star with 
its circumstellar disk as a companion, which was thought as  most likely cause for its long 
outburst of about 40 days, as seen for other Be-binaries. 
Optical observations of Cepheus X-4 subsequently confirmed 
it as  a Be-binary system and estimated its location  at a distance of $3.8 \pm 0.6$ kpc \citep{bonnet98}.
But this distance estimate was later challenged by \citet{riquelme12} who proposed a 
distance of either 7.9 or 5.9 kpc according to whether the stellar
type of the companion is a B1 or B2 star, respectively. The
distance of Cepheus X-4 was later reported by Gaia at $10.2^{+2.2}_{-1.6}$ kpc \citep{malac20}. 
Luminosity dependent changes in the pulse profile of Cepheus X-4 were studied 
during declining phase of 1997 outburst \citep{mukerjee00} by combining observations by
Indian X-ray Astronomy  Experiment \citep{agrawal97} along with RXTE \citep{roth98}. 
The  RXTE observed another outburst in 2002 and re-established  its cyclotron resonance
feature corresponding to  its centroid energy at $30.7 \pm 1.8$ keV (as established earlier 
by Ginga) which did not show a significant dependence on X-ray luminosity, although the continuum 
became harder with increasing source luminosity \citep{mcbride07}. 
The source went into outburst again in 2014 and was observed with the Nuclear 
Spectroscopic Telescope Array (NuSTAR), \citep{harrison13} and  Suzaku \citep{mitsuda07}. 
Results obtained from Suzaku observation of the 2014 outburst,  overlapped with NuSTAR second 
observation on July 1--2, 2014, which detected additional absorption  feature at $\approx45$ keV in the phase-resolved 
spectra of the pulsar, which was identified  as the first harmonic of the fundamental cyclotron 
line detected at $\approx28$ keV \citep{jaiswal15}. 
The source spectra obtained from NuSTAR observations of 2014 were well fitted by Fermi-Dirac cutoff (FD-cutoff)  
model along with iron emission line and cyclotron absorption feature was clearly detected in both the observations at  $30.39^{+0.17}_{-0.14}$ keV and $29.42^{+0.27}_{-0.24}$ keV,
respectively. Although,  the averaged source luminosity was different by a factor of about 3 between  these
two observations, however  it only showed marginal variation in its centroid energy \citep{furst15}.
Using same observations of NuSTAR of 2014 outburst \citep{vybornov17} reported that  spectrum of Cepheus X-4 
showed two cyclotron resonance scattering features, the fundamental line at $\approx 30$ keV and its harmonic
at $\approx 55$ keV. They also showed  that the energy of the fundamental cyclotron absorption feature increases 
and the continuum becomes harder with increasing X-ray luminosity. Pulse phase resolved spectroscopic  
studies of Cepheus X-4 were conducted by \citet{bhargava19} at these two different intensities of the source 
using same two  observations of NuSTAR of 2014 outburst. It was found that the observed cyclotron 
line profile of Cepheus X-4  was  of asymmetric shape for phase averaged spectrum. However, for phase resolved 
spectra, a  single symmetric cyclotron profile fitted the data well.  
The spectral continuum and  parameters of the cyclotron line, showed  some 
variations with respect to the pulse phase only within a limited pulse-phase \citep{bhargava19}.

Motivated with further studies and investigations on timing and spectroscopic  properties  
of Cepheus X-4 during its latest  2018 outburst,  AstroSat  observed the source,  
on July 2 and again on July 8, during the declining phase of the outburst  at 
two different source luminosities.  Results obtained  from these studies  
on timing and spectral properties of Cepheus X-4  at two different source luminosity 
during 2018 outburst are presented in this paper. The rest of the paper is organized such that  
Section 2 describes the observations and data analysis including software tools used, 
Section 3 describes the results and then   implications  of these results are discussed 
in Section 4. Finally, the conclusions drawn  from this study are summarized in Section 5.

\begin{figure}
\centerline{\includegraphics[scale=0.55,angle=-90]{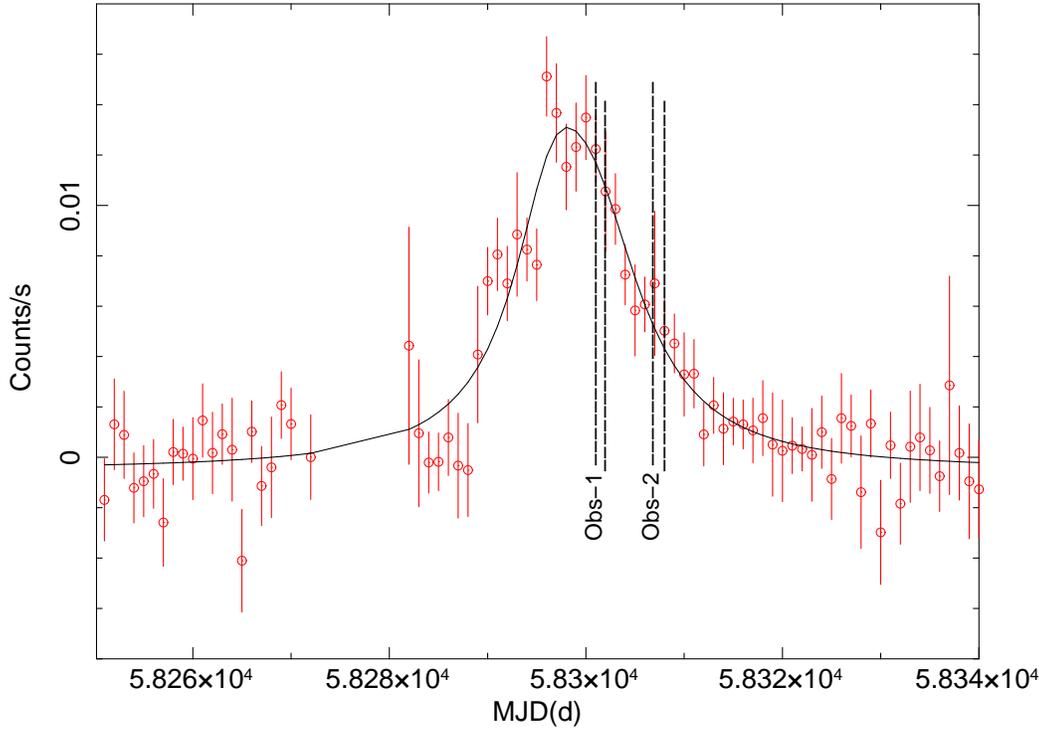}}
\caption{Intensity variation in 15--50 keV energy band  of Cepheus X-4 as observed by Swift-BAT 
during 2018 outburst is shown by data points while the continuous  line 
shows a fitted curve with a combination of two Lorentzian. Data points with very large error bars were
removed.  The start and end times for  two AstroSat observation intervals  are marked 
by  vertical dashed lines.}

\label{batfit}
\end{figure}
%
%

\section{Observations and Data Analysis }

Cepheus X-4 had another long outburst during  June--July, 2018 as
its latest event, as seen in the Swift-BAT intensity curve shown in Figure~\ref{batfit}.
The peak of the outburst  was  approximately determined by fitting a combination of
two Lorentzian to the Swift-BAT intensity curve.  The peak  of  the outburst occurred  
at MJD 58298.37 when the source intensity was about 60 mCrab 
(Swift-BAT 0.01308 cts cm$^{-2}$ s$^{-1}$) in the 15--50 keV energy band as determined from the 
fitted model.  AstroSat  observed the source  twice during its declining phase of 
the 2018 outburst under target of opportunity observation (TOO).  First observation was 
made during  July 2--3, 2018  between MJD 58301.61850 and 58302.56154 
with an Obsid T02\_055T01\_9000002206. This was followed by the second  observation 
during July 8--9, 2018 between MJD 58307.40211 and 58308.60036 with an  Obsid T02\_057T01\_9000002212.  
The start and end times of these observation intervals  are marked by  consecutive dashed vertical lines 
on Swift-BAT light curve
in Figure~\ref{batfit}. The source intensity  during these two observations were  about 52 mCrab and 27 mCrab,
respectively in the Swift-BAT energy band.  The  Swift-BAT light curve was obtained from 
Archive\footnote{https://swift.gsfc.nasa.gov/results/transients/index.html} and 
relevant descriptions about these  light curves are given by \citet{krimm13}.
The outburst of 2018  and the earlier outburst of 2014,
lasted for about the same duration of $\approx 45$ days but peak intensity of
the 2018 outburst $\approx 60$ mCrab (Figure~\ref{batfit}), was lower compared to $\approx 80$ mCrab 
during 2014 in the  15--50 keV energy band of the Swift-BAT \citep{furst15}. AstroSat observed the source 
during the outburst about 3-days and 9-days after attaining its peak luminosity, which is about 87\%  and 45\% of 
its  observed peak luminosity.

During the two AstroSat observations of Cepheus X-4, the Large Area X-ray Proportional
Counter (LAXPC) was used as the prime instrument for data  acquisition.  The LAXPC and 
its capabilities are  well described by \citet{antia17} and can be referred for details. 
During these observations only LAXPC20 was working nominally and was operated in its 
event-analysis mode. As LAXPC10 was operated at a very low gain hence data were not 
utilized for our studies.  An average source count-rate of 158 cts s$^{-1}$  and 
77 cts s$^{-1}$  were detected  by LAXPC20 in the 3--60 keV energy band.  
Total effective exposure for two observations of LAXPC were 44~ks and 54~ks respectively. 
The Soft X-ray Telescope (SXT) was operated in its photon counting mode during both observations,
with an effective exposure of about 20~ks and  29~ks. 
An average  count-rate of $2$ cts s$^{-1}$ and 
$1$ cts s$^{-1}$, were detected in SXT during two observations respectively in the 0.5--8 keV energy band.  
The SXT payload and its read-out modes are described by \citet{singh16}.  
The data from the AstroSat SXT and LAXPC20 were combined for detailed 
timing  and  spectroscopic studies in 0.3--60 keV energy band.  AstroSat data of these 
TOO proposals are available  in the  public domain at the AstroSat 
Data Archive\footnote{https://astrobrowse.issdc.gov.in/astro\_archive/archive/Home.jsp}.

The pipeline software version laxpcsoftv3.4\footnote{https://www.tifr.res.in/$\sim${}astrosat\_laxpc/LaxpcSoft.html}
was used for the  analysis of LAXPC data.  The pipeline software  runs by taking the required  input files from 
its  well defined directory structure which includes level-1 data files, calibration and background files 
for generation of level-2 data products. These included light curves  in the desired energy band and  its full 
range of spectrum.  The required background files are supplied along with the
software. The software itself offers guidance for selection of a suitable
background as well as a response file which may be used for further analysis.  
The background observation of August, 2018 was used for this analysis 
to estimate the contribution from the source.  The background observation of  
July 2018 as recommended  for use by the pipeline software was avoided as it posed problems 
particularly at energies of 30 keV and above, as the  model under-predicted the background 
counts affecting source contributions. The July 2018 background observations was most likely affected 
due to occurrence of a geomagnetic storm, hence was not considered.   
The pipeline software was executed with  default values of all the input parameters.  
We have considered all the events from main anodes from each layer of LAXPC20. 
Appropriate energy  channels were selected  during the software run  for extracting light curves for 
a required energy band. Gain corrections are applied automatically by the pipeline software itself 
to account for any shift in the gain between the source and background, based on calibration data. 
Thus the pipeline generates calibrated spectrum and background subtracted light curves of the source
within its selected energy band for further analysis of the data using other astronomical software
as mentioned later.

The pipeline analysis software specifically designed for SXT was used. 
The latest version of the software AS1SXTLevel2-1.4b, available at its payload operation center   
site\footnote{https://www.tifr.res.in/$\sim${}astrosat\_sxt/index.html} was used.
The software includes other files such as spectral response,
auxiliary response and background files, those are required for spectral data analysis.
The SXT calibration files are accessed automatically by the software from its well organized directory structure   
during its execution. The SXT level-1 data along with  House Keeping and calibration files are taken 
as input by the analysis software and produce cleaned and calibrated events lists.  
The ultimate data products such as spectrum, light curve and image were 
produced by the pipeline software, by filtering and screening of the events, for a set of
default parameters.  A circular region of interest of 6 arc minutes radius was 
considered with respect to the source center, for extraction of photon events for this analysis,  
covering all the grades between 0--12. We thus generated spectrum and light curves in different 
energy bands covering 0.3--8.0 keV. The  version 2.4c of the  {\tt Xselect} tool
was used for screening of the data. Timing analysis was performed after delay time corrections 
applied to all the events with respect to the solar system barycenter for both SXT and LAXPC data,
using a tool {\tt as1bary}\footnote{https://astrosat-ssc.iucaa.in/?q=data\_and\_analysis},
developed by the AstroSat science support cell.

Astronomical software HEASoft version 6.14 was employed for the higher level 
analysis of AstroSat data.  The timing analysis was done using software tools 
{\tt lcurve}, {\tt efsearch} and {\tt efold} etc. of  Xronos 5.22 package.  
The spectral data were analyzed using  Xspec software version 12.8.1.  
The standard spectral models supplied by the Xspec software were used. 
Additionally, a Fermi-Dirac cuttoff model was defined and used in combination 
with other Xspec models which are described below.  All these results are presented 
in the following section.

%
\begin{figure}
\centerline{\includegraphics[scale=0.36,angle=-90]{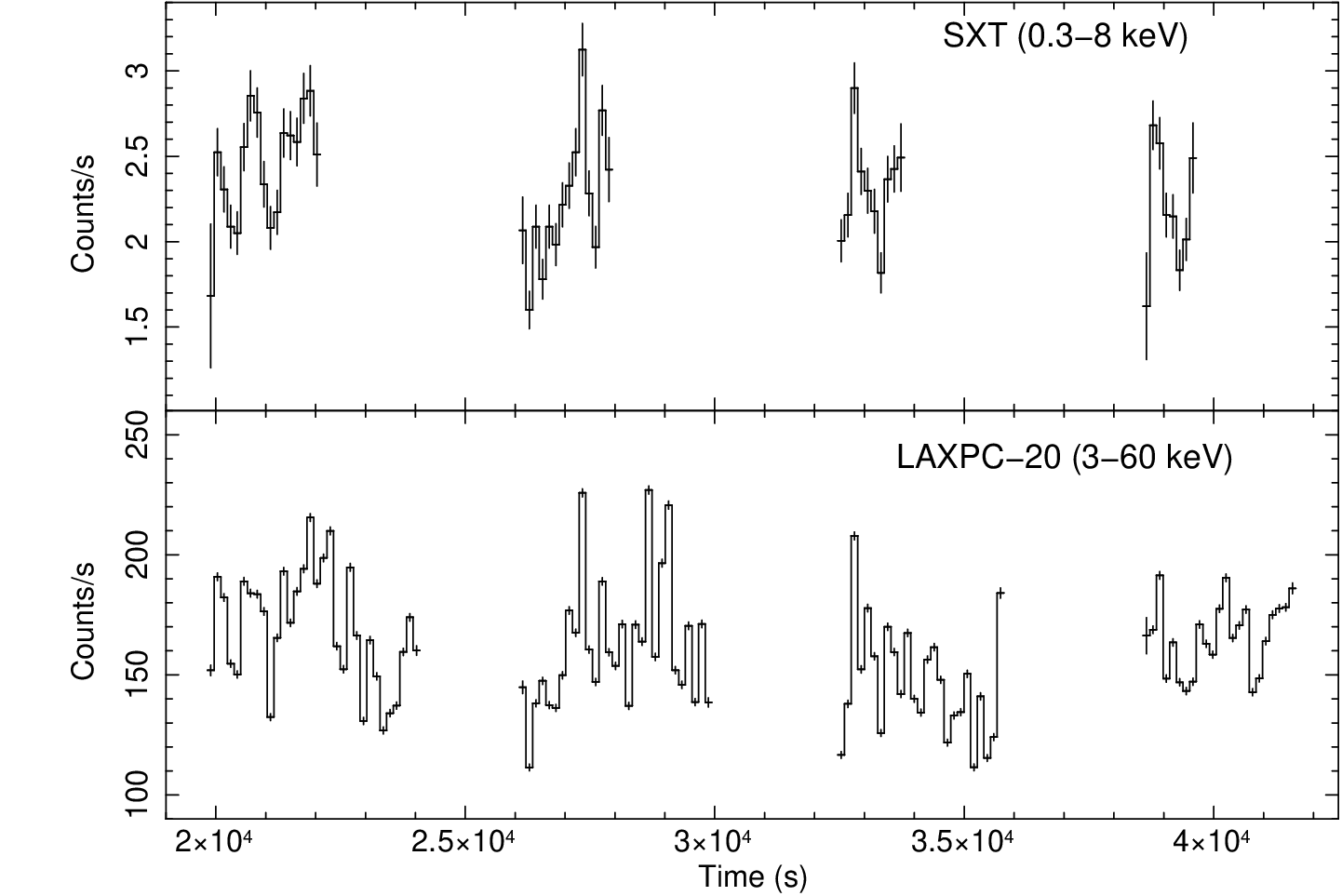}
\qquad\includegraphics[scale=0.36,angle=-90]{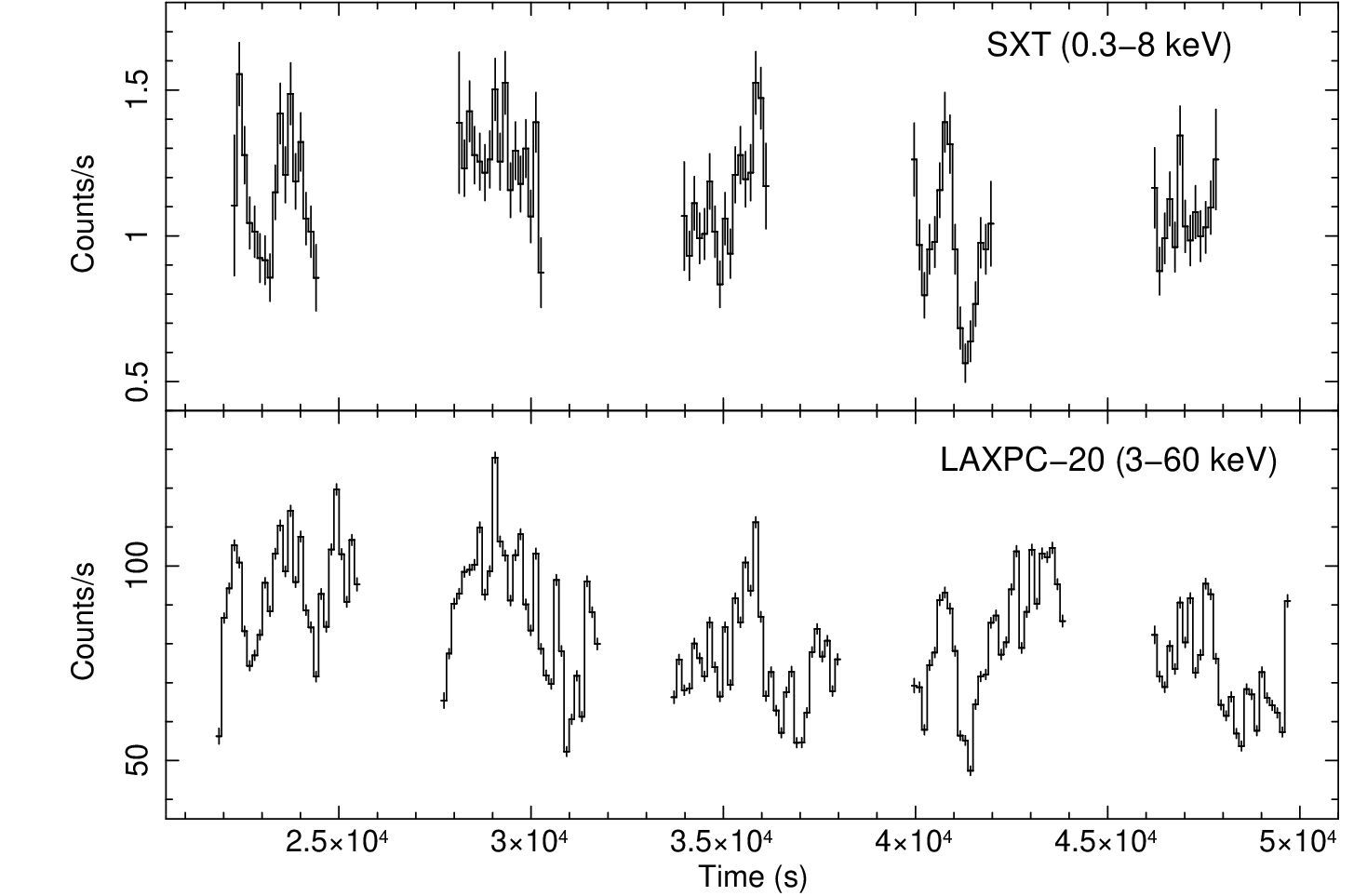}}
\caption{A common time segment of light curves  derived from SXT in 0.3--8.0 keV and 
LAXPC20 in 3--60 keV energy band  with 133 s binning are shown.  The light curve from July 2--3, 2018 
observation corresponds to a start time at MJD 58301.61850  (left panel) and  
from July 8--9, 2018 observation at MJD 58307.40211 (right panel) are displayed.}
\label{lc2}
\end{figure}
%

%
\section{Results} 
\subsection{X-ray Light curve and Pulse Period determination}
 
Light curves of Cepheus X-4 in 3--60 keV energy band from LAXPC20 with 1~s  time bin 
showed regular and strong 66 s pulsations along  with variation in its intensity. 
A  time segment of X-ray light curves after applying the solar system barycentric time correction, 
for  both  SXT and LAXPC20 with 133 s binning are shown  from the observations of 
July 2--3, 2018  (Figure~\ref{lc2}-left panel) and of July 8--9, 2018 (Figure~\ref{lc2}-right panel). 
Start time for these two observations are at MJD 58301.61850 and
MJD 58307.40211, respectively. 
The binning of 133 s, which is twice the pulse period was applied to the light curve to detect source flaring and 
dipping activities by suppressing  observed  intensity variations due to  spin-pulses and pulse to pulse variations 
in the high time resolution light curve and to  improve on statistical fluctuations.
Sudden flaring and dips were clearly  seen in the light curves of SXT and LAXPC20 
as can be seen in the Figure~\ref{lc2} along with its usual intensity variations during  
AstroSat observations.   These variations are typical property of a Be-binary pulsar 
and is also observed in the case of Cepheus X-4 during AstroSat observations.\\
 
Using photon arrival time delay corrected light curves, first, average spin-period of the pulsar 
was determined using epoch folding technique
for each of these two AstroSat observations along with its  estimated  
1-$\sigma$ error as  $66.351\pm0.008$ s and $66.353\pm0.008$ s respectively. 
The uncertainties are the estimated 1-$\sigma$  errors from Gaussian function fit  to the  
values of calculated $\chi^2$ versus a  range of period curve, where  best-period is chosen  
corresponding to  the Gaussian peak \citep{pravdo01}. The estimated errors so derived  are  
rather an over estimate of the measurement errors. 
From  the difference of these spin periods, an  average  spin-down rate of the order 
of $\approx 1\times10^{-12}$ Hz s$^{-1}$  was determined 
for the purpose of an initial information. 
Now, the  pulsar spin period  was  determined with higher precision, at an epoch corresponding to 
the start of the observation  at $t=t_0$ along with its time derivative independently for two
observations, by applying the phase correction technique using
\begin{equation}
	\phi(t)=\phi_0+\nu_0(t-t_0)+\dot\nu{(t-t_0)^2\over 2},
	\label{eq:phase}
\end{equation}
where $\phi$ is the phase which covers the range 0--1 of the period, $\phi_0$ is the phase, 
$\nu_0$ is the spin frequency and $\dot\nu$ is its time derivative,  all corresponding to the start time $t_0$.
The fit was performed by fitting a periodic signal and considering reasonable numbers of harmonics of the basic
spin frequency using 
\begin{equation}
	c(t)=c_0+\sum_{k=1}^N\Big(a_k\sin(2\pi k\phi)+b_k\cos(2\pi k\phi)\Big),
	\label{eq:fit}
\end{equation}
where $c(t)$ is the observed count rate, $N$ is the number of harmonics included in the fit
and $c_0,a_k,b_k$ are the coefficients fitted, apart from $\nu_0$ and $\dot\nu$.
The light curve with a time resolution of 1 s was obtained for the analysis. 
The best fitted values for spin-frequency $\nu_0$ and its derivative
$\dot\nu$  were determined by minimizing $\chi^2$  deviation of the
light curve from the model defined by Eq.~\ref{eq:fit}.  Different number of harmonics $N$ were tried and
it was found that $N=20$ is adequate to define the signal appropriately for the analysis.  
Monte Carlo simulations were done to determine errors in the fitted values of $\nu$ and $\dot\nu$. 
The simulations were performed by perturbing $c(t)$ and repeating the process for 4000 different 
realization of noise and thus 90\% confidence limits for the parameters were determined.
The program was also used to extract data for pulse-profile generation and to 
define good time intervals for different phase intervals for extraction of spectra using 
the fitted values of parameters, for studies of phase resolved spectroscopy.
The fitted value of the spin-period of the pulsar and its spin-up rate were  determined as  
$P=66.35080\pm0.00014$ s and  $\dot\nu=(-2.1\pm0.8)\times10^{-12}$ Hz s$^{-1}$
corresponding to MJD 58301.61850 for the first AstroSat observation.  Whereas for 
the second observation, it was determined as $P=66.35290 \pm 0.00017$  and 
$\dot\nu=(-1.6\pm0.8)\times10^{-12}$ Hz s$^{-1}$ at an epoch MJD 58307.40211. 
The quoted error denotes 90\% confidence limit for both the observations.
The effect of orbital motion on these parameters could not be determined, as orbit of   
this binary is still unknown. 

%
\begin{figure}
\centerline{\includegraphics[scale=0.35,angle=-90]{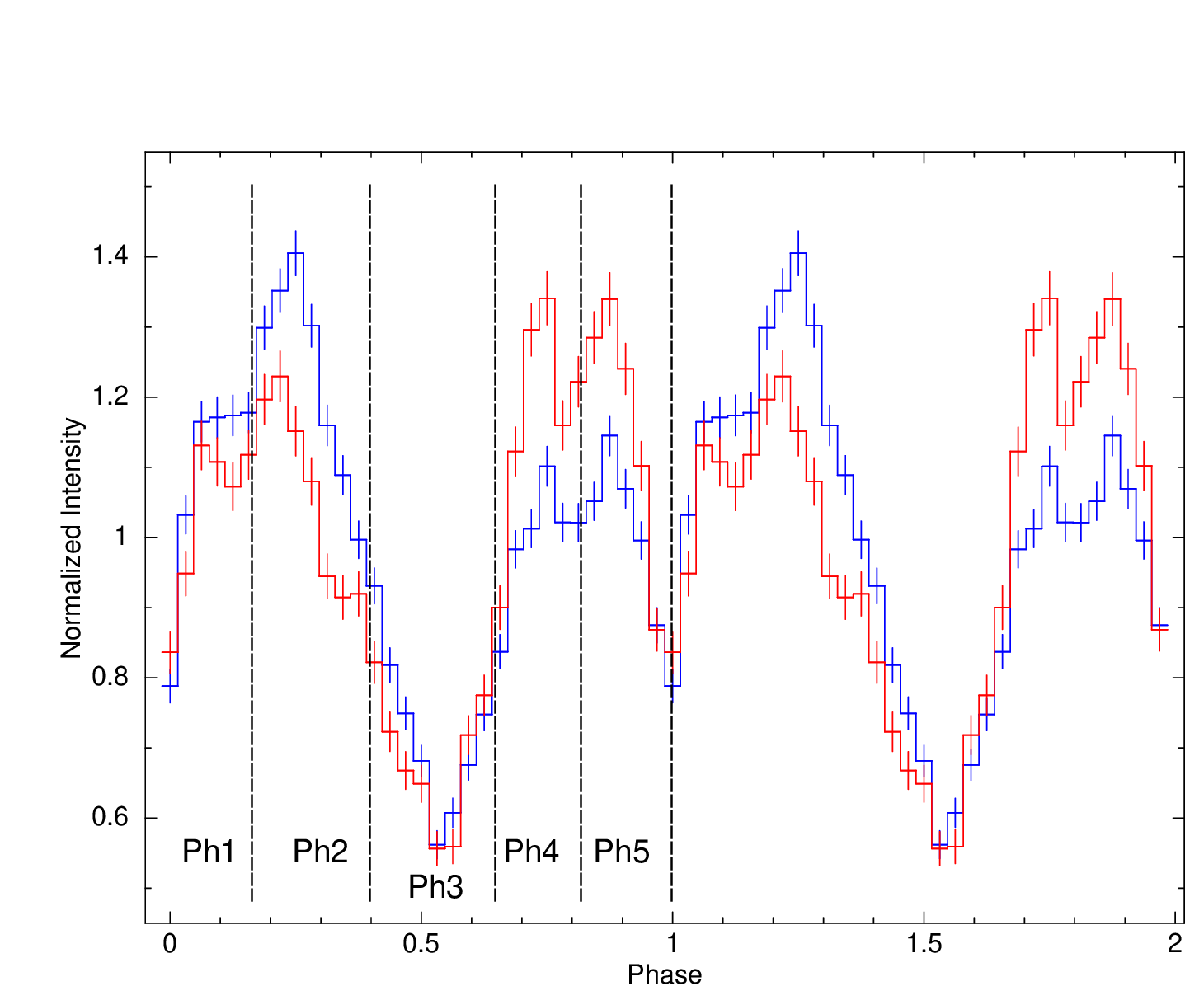}
\qquad\includegraphics[scale=0.35,angle=-90]{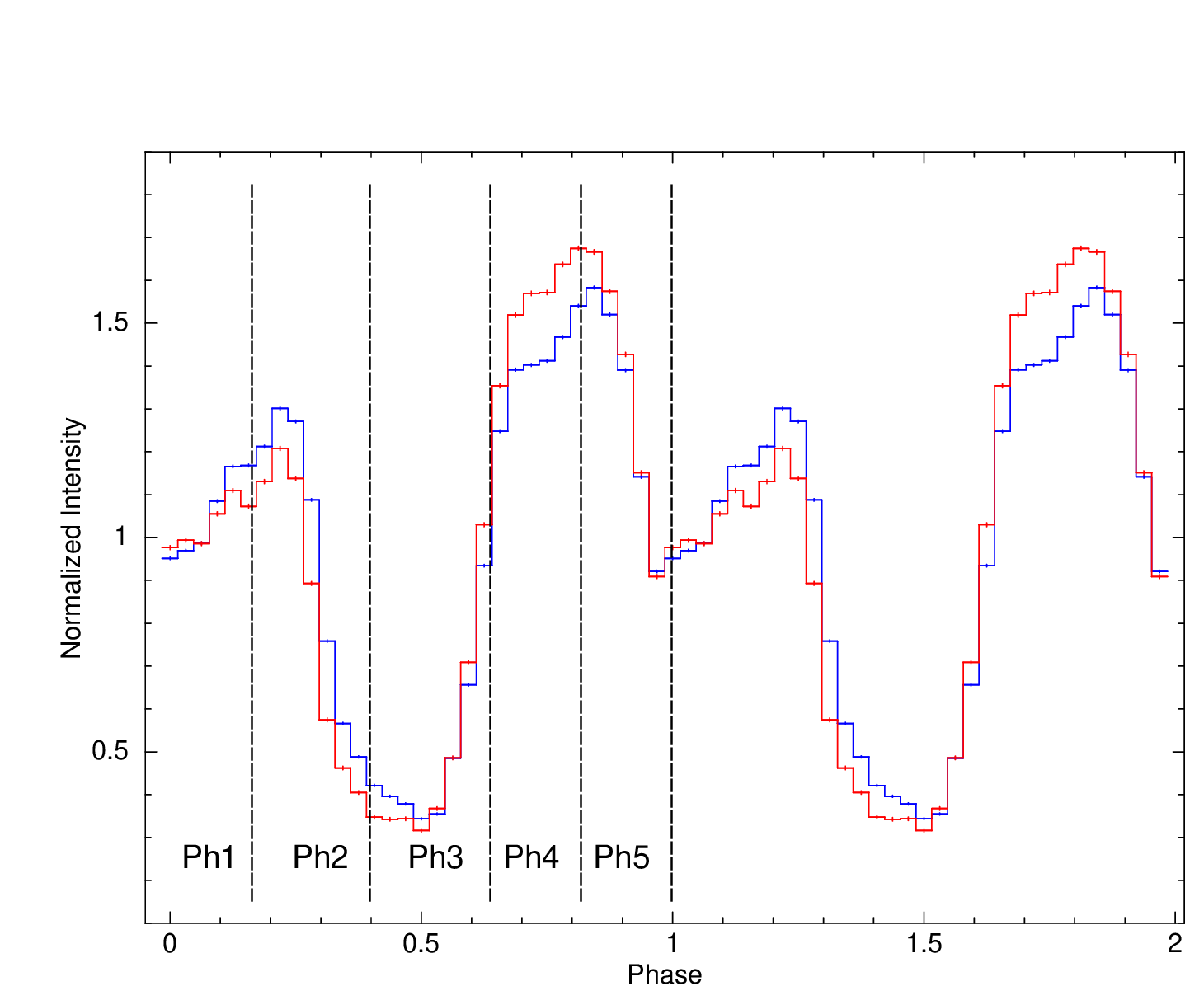}}
\caption{Pulse-profiles derived from SXT in 0.3--8 keV energy band (left panel) and LAXPC in 3--60 keV 
energy band (right panel) from two AstroSat observations are shown with an overlap. The blue colored profile 
is derived from July 2--3, 2018 and red colored profile from July 8--9, 2018 observation. 
The 5 different phase bins used in the phase resolved study are marked in the figure.}
\label{pulsea}
\end{figure}
%
\begin{figure}
\centerline{\includegraphics[width=10cm,height=8cm,angle=-90]{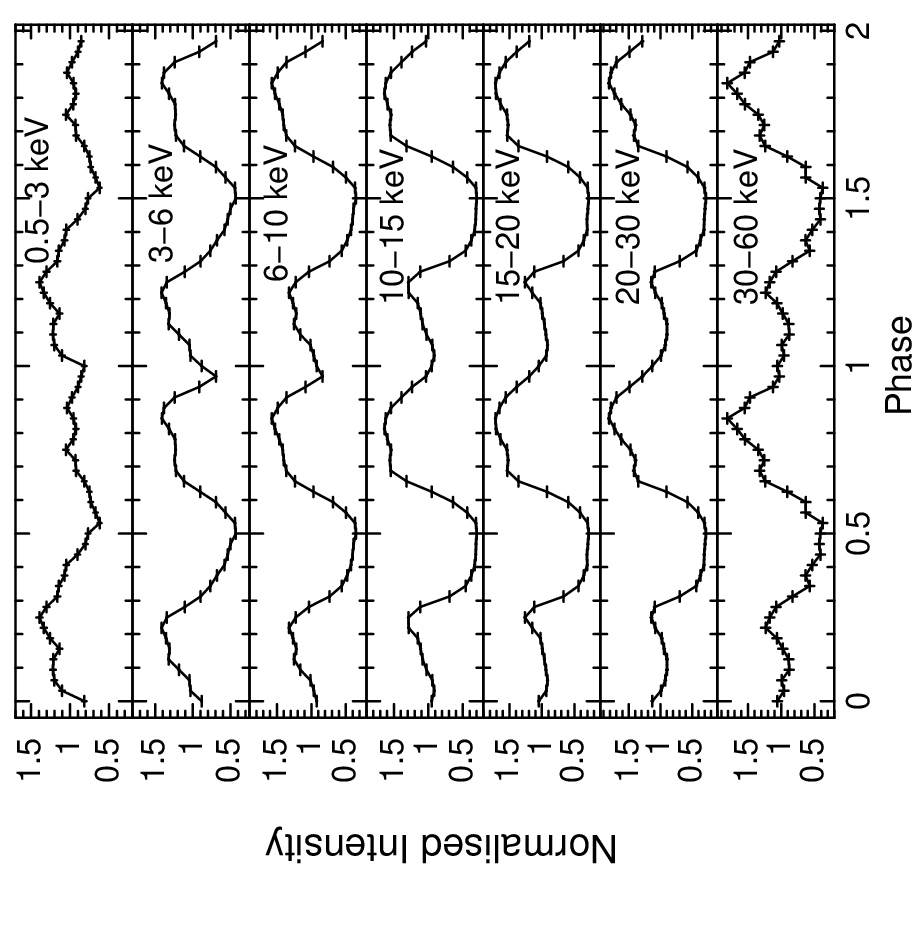}
\qquad\includegraphics[width=10cm,height=8cm,angle=-90]{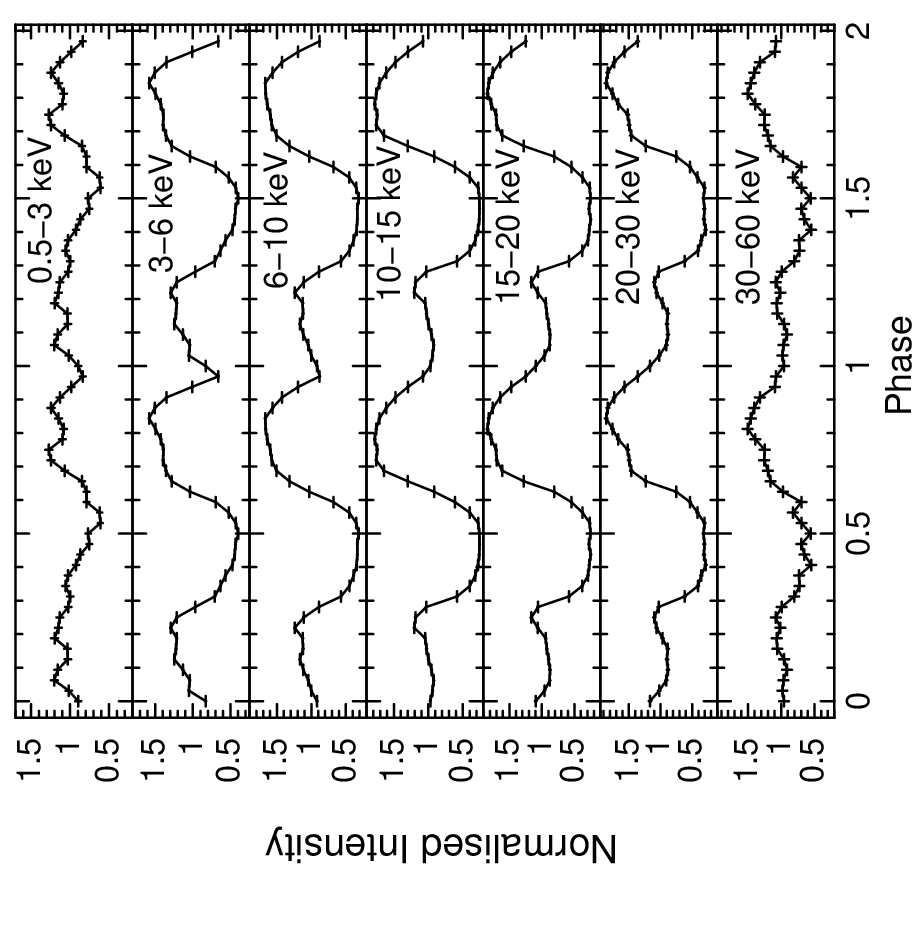}}
\caption{Folded pulse profiles  derived from  SXT and LAXPC20 at different energy bands 
covering 0.5--60 keV during the two AstroSat observations.  The left panel shows the result 
from the first observation (July 2--3, 2018), while the right panels shows that from 
the second observation (July 8--9, 2018).}
\label{pulseb}
\end{figure}
%

\subsection{Pulse profiles and Pulse fraction Measurements}

Light curves were extracted  corresponding to  different energy bands and  barycentric corrections 
were applied. Respective pulse profiles were derived by folding  corresponding  light curves  with its 
derived spin-period and its derivative for the two observations.  Small phase-shifts were applied  to 
the corresponding  start of the epochs to match the phase of these pulse profiles for the two observations.
Pulse profiles were derived for observation-1, using 3--60 keV light curve of LAXPC20 first, such that the
minimum of the profile corresponds to pulse phase 0.5. Same  reference epoch was used for folding  0.3--8 keV 
SXT light curve  for observation-1. Similar procedure was followed for observation-2.  
Two cycles of all such pulse-profiles  are shown in Figures for clarity. 
The pulse profiles  showed mainly two prominent pulses, the pulse between the  phase 0.5--1.0 (1.5--2.0)
covers the primary-pulse and phase between 0.0--0.5 (1.0--1.5) the secondary-pulse, as can be seen 
in Figure~\ref{pulsea}.  
Folded pulse-profiles showed variation  at different energy with respect to  the pulsar spin-phase. 
The average pulse-profile for both observations covering energy band between  0.3--8 keV of SXT and 
between 3--60 keV of LAXPC20 
are overlayed in the Figure~\ref{pulsea} for relative comparison. The relative
intensities of primary pulse compared to its secondary showed increase with decrease in source luminosity
at both energy bands  0.3--8 keV (left panel) and 3--60 keV (right panel) as can be seen in Figure~\ref{pulsea}.
For detailed comparison, we derived pulse profiles at 7-different energy bands  between  0.5--60 keV from 
two AstroSat observations and results are shown in Figure~\ref{pulseb} for observation-1 (left panel) 
and  observation-2 (right panel).  The pulse profile between  0.5--3 keV energy band was derived from SXT data, where 
both the pulses, primary and
secondary showed multiple structures at lower energy bands and are most prominent  between 0.5--3.0 keV. 
Where as for higher energy bands, it showed two pulses with change in its relative intensities 
depending on energy and source luminosity. The relative  intensity of primary pulse increases with 
energy and the secondary pulse becomes gradually narrower. Moreover, the sharp narrow dip at lower 
energies  changes to a  smoother  and broader shape  connecting primary and secondary pulses as can be  seen during 
both the observations.  It is also noticed that the presence of multiple structures and absorption dips 
of both primary and secondary  pulses as seen at  lower energies,  gradually  merged into a 
double-pulse  profile with gradual increase in the energy and then becomes asymmetric and broad 
with smoothening of the dip joining primary and secondary  at higher energies. We also noticed 
that  for observation-2 at lower luminosity the primary pulse becomes relatively stronger than the 
secondary pulse in contrast to observation-1 for higher  source luminosity.

\begin{figure}
\centerline{\includegraphics[scale=0.45,angle=-90]{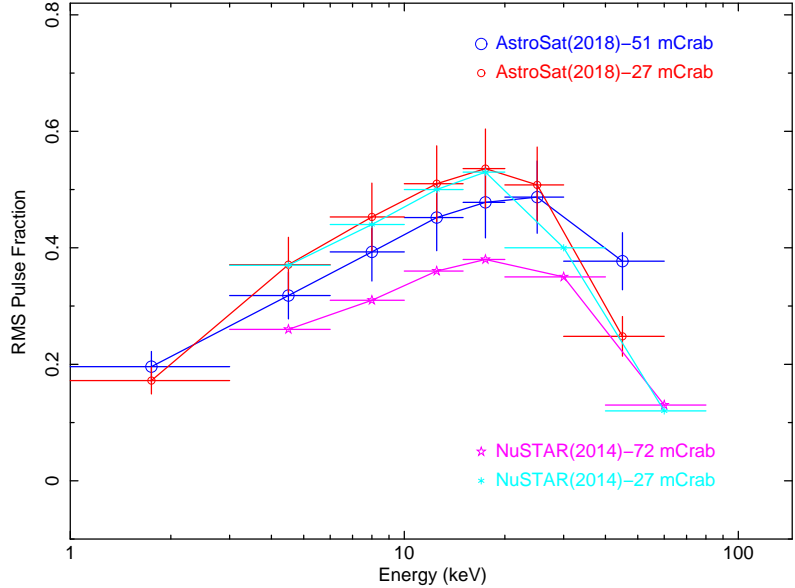}}
\caption{RMS Pulse fraction variation with  energy and the source intensity  during
2 outbursts of Cepheus X-4  as observed by different observatories 
NuSTAR (2014) and AstroSat (2018) are shown.}
\label{pfrac}
\end{figure}

Pulse profiles showed  remarkable variability in shape and its 
pulse-fraction with respect to energy and source luminosity as seen in Figure~\ref{pulseb} and 
Figure~\ref{pfrac} respectively.  
The RMS pulse fraction varied from  about $32 \pm 4$\%  at 3--6 keV,  
$48 \pm 6$\% at 15--20 keV and $38 \pm 5$\% at 30--60 keV  during observation-1.   
The corresponding measurements for observation-2 showed  $37 \pm 5 $\%,
$54 \pm 7 $\%  and $25 \pm 3$\%, respectively. 
Pulse-profiles derived from recent AstroSat observations  are though 
similar in shape but structures can be seen more clearly, as compared to those reported from 
November 28, 1996 RXTE observation between 2--60 keV during its earlier 
outburst \citep{wilson99}. 
RMS pulse fraction  measurements from  2 outbursts of Cepheus X-4, observed by  
NuSTAR in 2014  and with AstroSat  in 2018, are  shown in Figure~\ref{pfrac}.
The pulse shape and pulse fractions measurements
of AstroSat observation-2 corresponding to  27 mCrab  Swift-BAT intensity in 15-50 keV energy band, 
showed consistency with the
NuSTAR observation at similar intensity during 2014 outburst \citep{bhargava19}.
These plots show that  the RMS pulse fraction varies with  both the energy and  the source intensity.
The pulse fraction measurements starting from highest intensity of source at 72 mCrab  showed gradual
increase as source intensity decrease to 52 mCrab and 27 mCrab attaining its peak-value around 20 keV.\\ 

\subsection{Spectrum and Cyclotron Resonance Scattering feature}
\begin{figure}
\centerline{\includegraphics[scale=0.36,angle=-90]{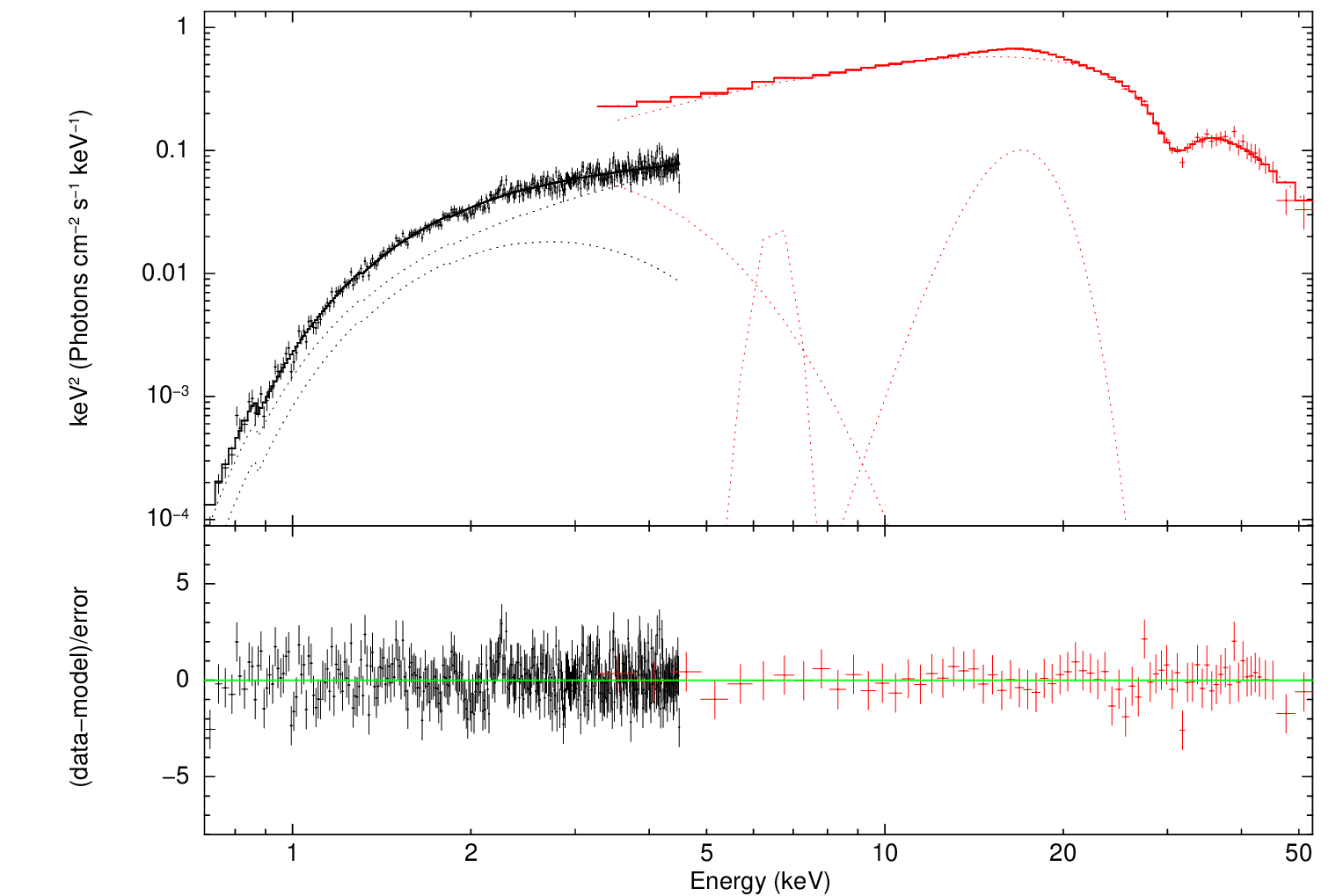}
\quad\includegraphics[scale=0.36,angle=-90]{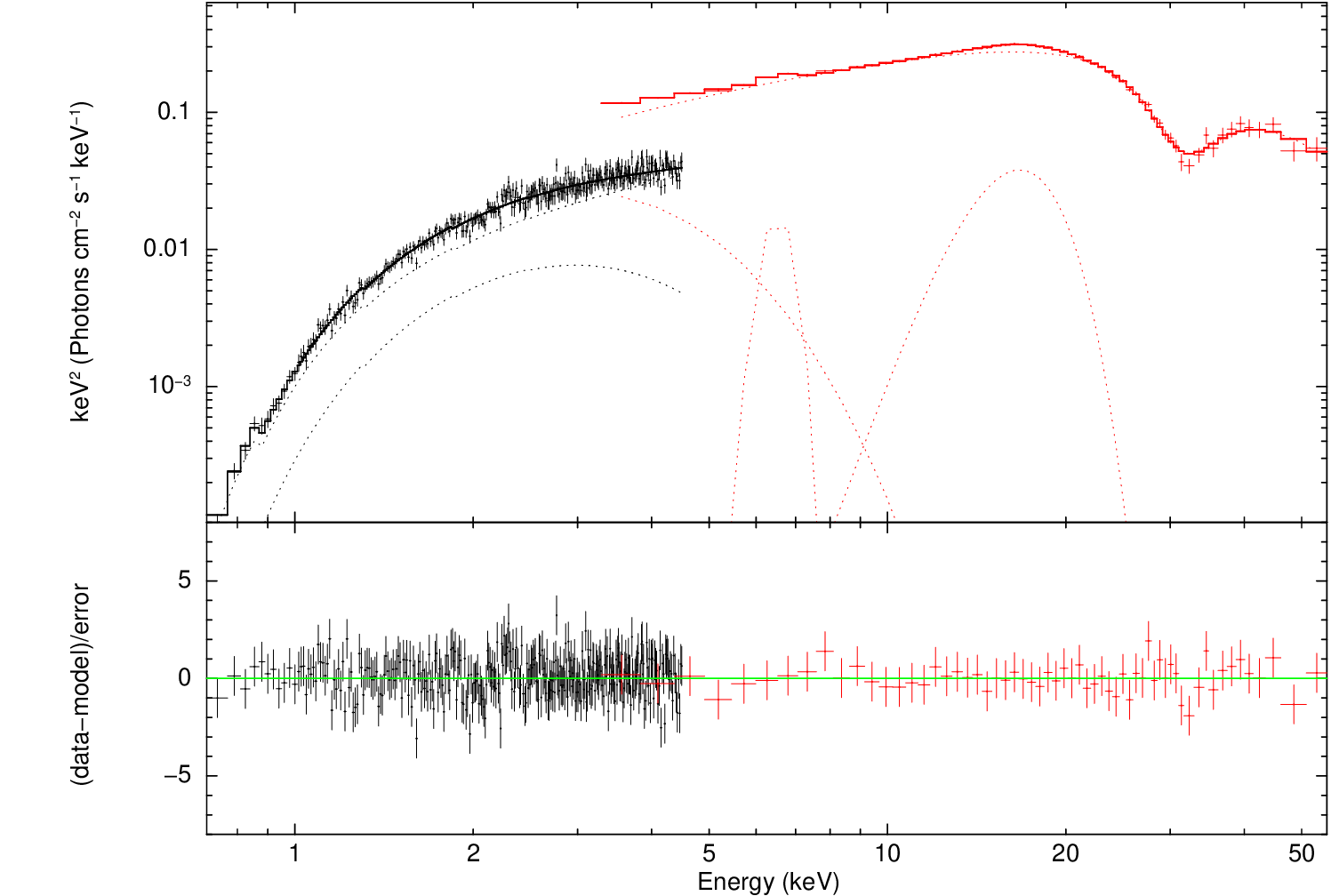}}
\caption{Spectrum  fitted with  combined model;  an absorbed power law with Fermi-Dirac
cut-off, along with a black-body, a cyclotron absorption line and an iron emission
line   for observation~1 (left panel) and observation~2 (right panel).  
The residue to the fit are shown below. An additional Gaussian line was introduced to model
16 keV feature in the pulsar spectrum.}
\label{spec1}
\end{figure}

\begin{figure}
\centerline{\includegraphics[scale=0.36,angle=-90]{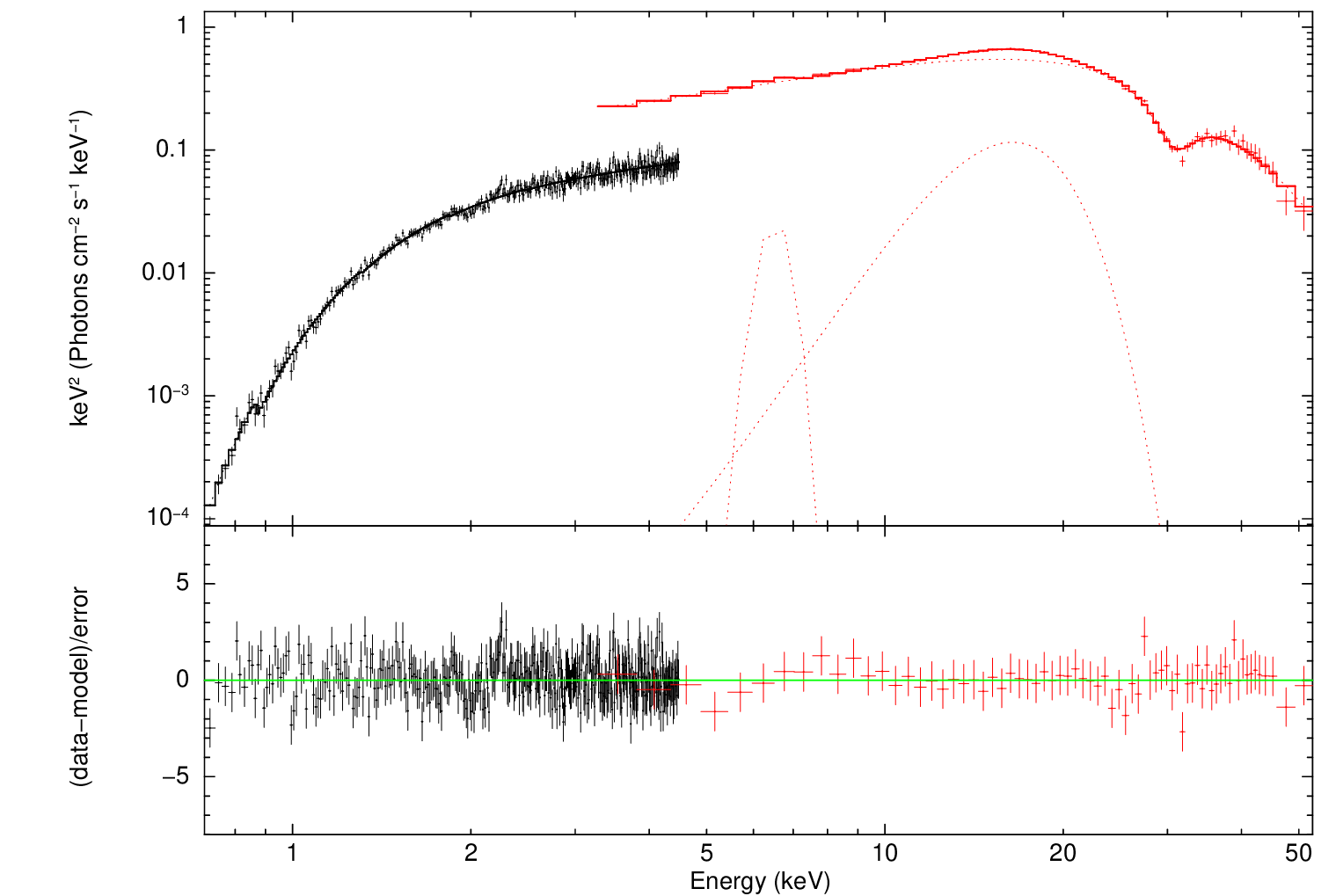}
\quad\includegraphics[scale=0.36,angle=-90]{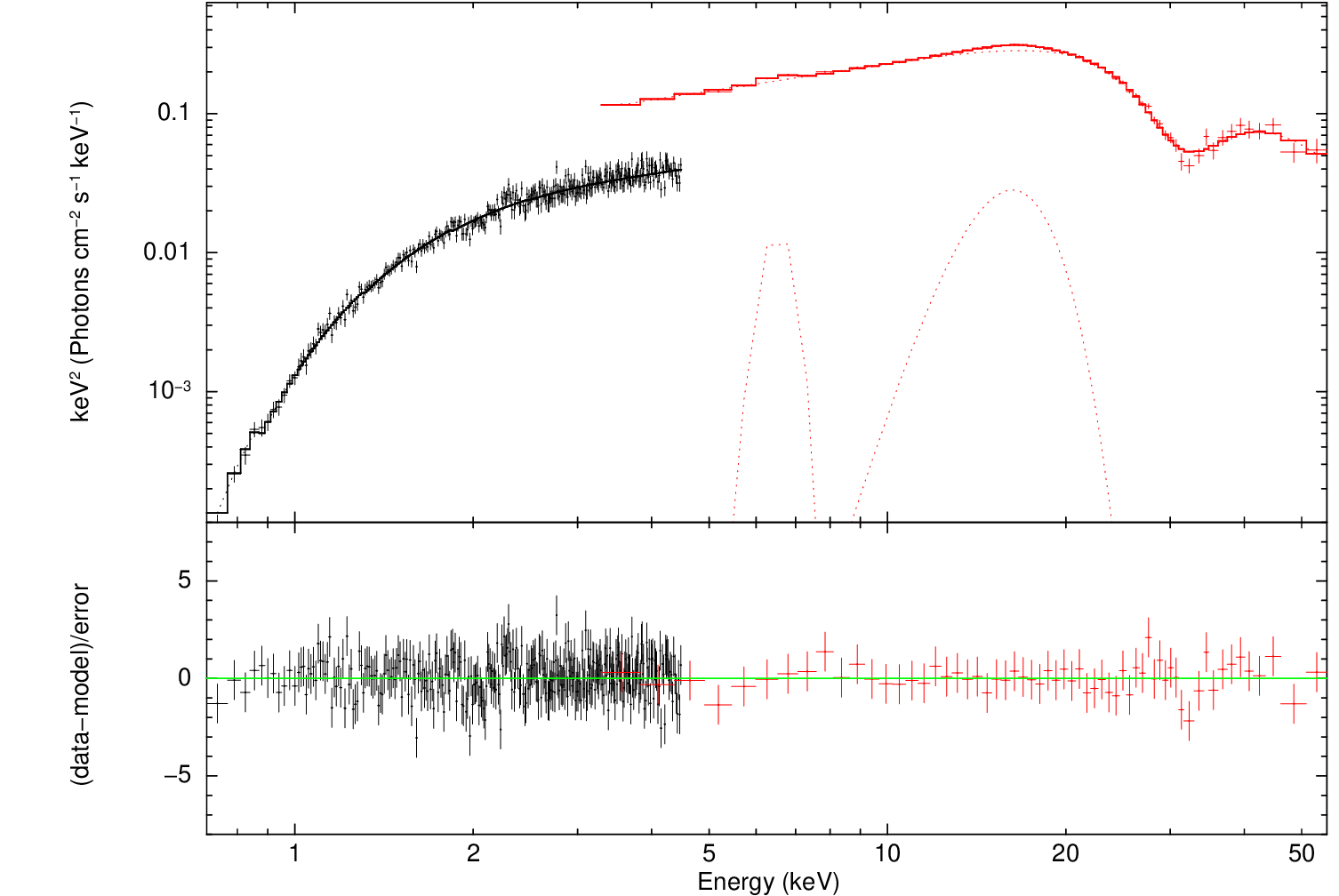}}
\caption{Spectrum  fitted with  combined model;  an absorbed thermal Comptonization 
(compTT) along with a cyclotron absorption line and an iron emission
line  for observation~1 (left panel) and observation~2 (right panel).  
The residue to the fit are shown below.  An additional Gaussian line was 
introduced to model 16 keV feature in the pulsar spectrum.}
\label{spec2}
\end{figure}

AstroSat observations enabled us to derive X-ray spectrum in  0.7--55 keV energy band  at 
two different source intensities  which
fitted reasonably  well using two continuum models. The first model was defined as an absorbed 
Fermi-Dirac cutoff (FD-cutoff) 
model \citep{tanaka86} along with a black-body, a Fe-emission line and a cyclotron absorption line 
as shown in Figure~\ref{spec1}.  
The compTT  model \citep{tit94} was used as the second model in combination with a Fe-emission line 
and  a cyclotron absorption line  as shown in Figure~\ref{spec2}. 
In most cases, the compTT-model itself defines  pulsar spectrum adequately well at lower energies,
without a black-body component \citep{doroshenko10, maitra13, jaiswal15, yoshida17, mukerjee20} unlike
cutoff-models combined with a black-body. However, in some cases black-body model was combined with compTT-model to
appropriately define the low energy spectrum such as for 1A 1118-61 \citep{suchy11}. For AstroSat
phase-averaged  spectra, addition of a black-body component with the compTT-model
changed the overall ${\chi^2}$ only by $\approx 1$ and resulted in  large errors  of the black-body parameters,
hence it did not offer any useful results.  We, therefore, used  a compTT-model for Cepheus X-4  without a back-body
component which defined its continuum  well covering the energy band 0.7--55 keV, like it was used earlier for
Suzaku observations \citep{jaiswal15}.
The fitted model parameters  along with estimated flux for two observations are given in Table~1.  
The relative normalization between the two instruments were modeled by introducing a constant term in the
combined model. Then by fixing the normalization of LAXPC20 to 1, the relative  normalization 
of the SXT was found to be 0.3 due to its fixed offset with respect to LAXPC20.
The SXT spectral data were restricted to the energy range within 0.7--4.6 keV  
to offer a good statistics to its spectral data in the range which helped in improving the overall spectral fit.
An overall systematic error of $2.5\%$ was introduced during  spectral fitting to account for uncertainties in
response matrices of both the instruments.  
The Cepheus X-4 spectrum  during two AstroSat observations shows significant change in its continuum as can be 
seen  by comparing spectral parameters as given in the  Table~1.  
A single photo-electric absorption (phabs) component is employed in both the models for two observations. 
The combined effect of absorption due to hydrogen column density  
and due to presence of matter surrounding the neutron star should result in relatively lower 
net absorption for lower source luminosity.
Therefore phabs is relatively lower for the second AstroSat observation of 
Cepheus X-4 as seen in the Table~1.  This is consistent with earlier  observed case of GX 1+4,  where  
corresponding to lower source luminosity in 2--60 keV energy band, the phabs were lower as can be 
observed  in Table~1 of \citet{galloway00}. 
Moreover, the photo-electric  absorption
is usually found relatively lower when  compTT-model is employed compared to  when high energy cutoff models 
combined with a black-body are used \citep{enoto14, mukerjee20}. 
This is also observed here for the case of Cepheus X-4. The compTT-model does reflect this change  appropriately 
as seen in GX 1+4 and also in the present case of Cepheus X-4 from Table~1. 
The Fe-emission line was prominently detected by LAXPC during both the 
AstroSat observations and its energy and width were fixed to appropriately model other features of  spectra. 

A residual feature around 30 keV, which is sometimes observed in the LAXPC
spectra  in many sources, \citep{antia21, beri21, jaiswal20, sridhar19, sreehari19}, 
was found to be very weak and barely detectable in  LAXPC spectra of Cepheus X-4. This is similar to the case of 
GRO J2058+42 \citep{mukerjee20}, and hence it would not pose any implications for measurements of the 
spectral continuum and cyclotron-line parameters for Cepheus X-4. 

Spectral data when fitted with a variety of continuum models, additional features 
were observed in the spectral residuals of many accretion powered X-ray pulsars. 
These features were observed either in emission between energy band  10--20 keV or more rarely 
in absorption between 8--10 keV \citep{coburn02}.  
It was subsequently argued and understood that such features may be caused more 
likely because of inadequacies of the continuum models rather than 
due to presence of cyclotron resonance features \citep{coburn02}. 
For example, this feature  was  observed as an emission-line between 10--20 keV
for Vela X-1 \citep{kreykenbohm02}, Her X-1 \citep{coburn02},  
Cepheus X-4 \citep{mcbride07} while as  an absorption-line between 8--10 keV 
for 4U 1907+09, 4U 1538-52 and 4U 0352+309, as can be seen in Figure~6 of \citet{coburn02}.
During the RXTE observations of Cepheus X-4, a similar feature was clearly seen   
and modeled using $\sim14$ keV Gaussian line \citep{mcbride07}.   
This is in contrast to the NuSTAR data where a broad feature with 6 keV line-width and centered 
around 12--13 keV was observed \citep{vybornov17} during its two observations of Cepheus X-4 during 2014 outburst. 
Interestingly, similar feature between 10--20 keV  was also observed  by \citet{furst15} (in their Figure-2(b)) 
during  studies of Cepheus X-4 using the same NuSTAR observations.    
An absorption feature around 19 keV was added in their model, to account for an asymmetry of 
the main absorption line profile. Most likely due to inclusion of 19 keV line, the observed feature in emission 
between 10--20 keV was also smoothed out during the  spectral fit. Similar approach was followed by  \citet{bhargava19}  
for their studies using the same  2014 NuSTAR data, however, no absorption line around 19 keV was introduced for 
their studies of phase resolved spectroscopy.  This is in contrast to the fact that such broad-feature was clearly 
seen in the residuals of phase-averaged spectra of two NuSTAR observation of 2014 as reported by \citet{vybornov17}. 
We find presence of 16 keV feature in the phase averaged spectra for both the
observations of AstroSat and hence a Gaussian component was included in the combined model as can be seen in the Table~1. 
We find 16 keV feature is relatively weaker for the observation-2 and has no significant  interference with 
the cyclotron line. For observation-1, the Gaussian width of the 16 keV feature was measured from the residuals of 
the fitted spectral model without it and then fixed its width for the overall 
spectral modeling, particularly for FD-cutoff model to reduce its interference with the cyclotron line measurements.  
The effect of 16 keV features on spectral continuum  was found to be in-significant for both the models 
for two observations, as all parameters values are found to be  consistent within their respective  error limits. 
The cyclotron-line width,  particularly for observation-1, is found to be significantly lower due to introduction of 16 keV
feature in the model. The cyclotron-line energy, however,  are either  
marginally affected or remains consistent like its optical-depth within their parameter error limits. 
The $E_{cut}$ parameter  for observation-2 showed wide variation with respect to observation-1 for FD-cutoff model, 
with or without the inclusion of 16 keV feature in the spectral model,  
which was not seen in all earlier observations by other instruments. 
Therefore, $E_{cut}$  parameter was frozen at a value as in observation-1 and then all other  parameter  were estimated 
for observation-2.

The derived  spectral parameters from FD-cutoff model showed that the source spectrum 
is relatively harder for higher source luminosity. Where as the black-body 
temperature is although  relatively lower at $0.62^{+0.04}_{-0.05}$ keV 
at higher source luminosity than  the temperature $0.69^{+0.04}_{-0.05}$ keV 
observed  at the lower luminosity but are within their error limits. This is similar to earlier    
observation of  Cepheus X-4 during 2014 where back-body temperatures   
were found to be at $0.92^{+0.02}_{-0.02}$ keV and 
at $1.04^{+0.03}_{-0.04}$ keV at two luminosities \citep{bhargava19}
but at higher values compared to 2018 observations. 
The compTT model could  successfully  define  spectrum  of many  
X-ray pulsars in high mass X-ray binaries for example Cepheus X-4 \citep{jaiswal15}, 
4U 1907+09 \citep{varun19} and GRO J2058+42 \citep{mol19, mukerjee20}.
Though, It is generally applied to  neutron star based low mass X-ray binaries  such as 
Z-type  and atoll sources having relatively lower  magnetic field ($<10^{12}$ G) \citep{ferinelli08}.  
The FD-cutoff model  estimated  black-body temperatures at two different luminosity of the source 
along with other spectral and cyclotron line parameters for both the AstroSat observations (Table~1). 
The Comptonization model on the other hand
enabled us to determine input photon  Wien-temperature of $0.42^{+0.04}_{-0.03}$ keV and $0.49 \pm 0.03$, 
the plasma temperature of $5.17^{+0.25}_{-0.17}$ keV and $6.92^{+0.62}_{-0.49}$ keV, along with its corresponding
plasma optical depth  of $6.05^{+0.26}_{-0.53}$  and $5.21^{+0.22}_{-0.30}$ respectively for  
the phase-averaged spectra.  
As bulk Comptonization occurs as per compTT model, in the innermost part of the transition-layer  
presumably within some extended region located away from the neutron star, while thermal Comptonization 
is dominant in the outer transition layer.  The Wien-temperature is, therefore, always found to be relatively lower 
than the neutron star surface black-body temperature  as it originates away from the neutron star surface 
at the outer transition layer \citep{ferinelli08}. Difference in these two temperatures were also observed earlier in 
spectral data of Cepheus X-4 fitted with compTT-model and FD-cutoff model combined with a black-body \citep[Table 1 of][]{jaiswal15}. 
The cyclotron line  parameters are found to be consistent well within the error limits of  the
two models. The cyclotron line parameters when compared for two different luminosity of the source
showed consistency in cyclotron line energy, however, line width and the line depth 
were found relatively larger at lower luminosity of the source.

%

\begin{figure}
\centerline{\includegraphics[scale=0.45,angle=-90]{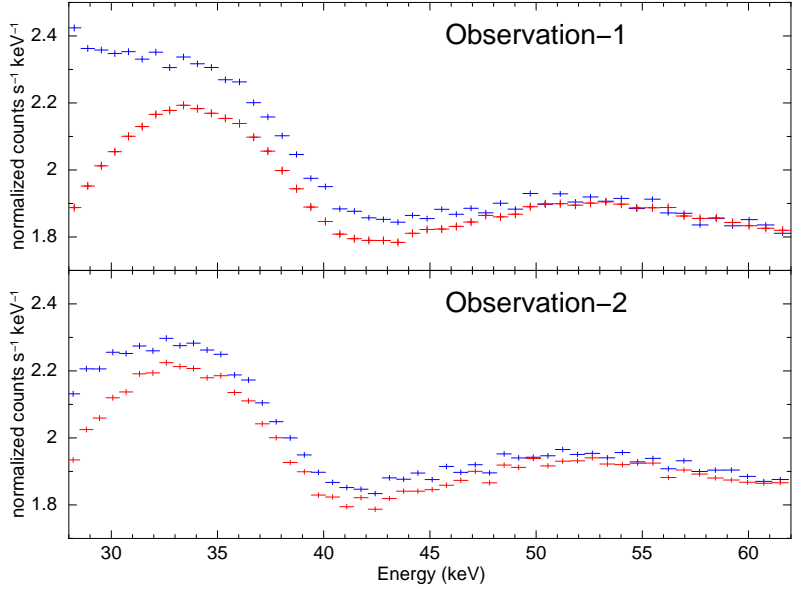}}
\caption{ Comparison of spectral counts of source plus background (Blue points) and background (Red points) of LAXPC20  
for observation~1 (top panel) and observation~2 (bottom panel).}  
\label{sbcomp}
\end{figure}

The detection of a second cyclotron absorption feature in the spectrum was reported 
earlier during 2014 outburst around 45 keV by \citet{jaiswal15} in Suzaku
observation, and  around 55 keV by 
\citet{vybornov17} in NuSTAR observation. AstroSat LAXPC20 is having effective area of about 1500 cm$^2$ at 
50 keV and 1300 cm$^2$ at 60 keV \citep{antia17} in comparison to about 150 cm$^2$ at 50 keV 
and 130 cm$^2$ at 60 keV  of NuSTAR \citep{harrison13}.  AstroSat, thus has superior 
effective area of an order of magnitude higher than the NuSTAR in these energies,
though uncertainties in LAXPC background are higher at these energies. We, therefore, 
made a comparison of spectral counts at two observed source intensities of AstroSat.  The 
combined source plus background and background  counts of LAXPC20  with $1\sigma$ 
errors on data  are  shown in Figure~\ref{sbcomp}. These  plots  show detection of source 
above the background spectrum until it  merges with the background around 52--55 keV.
We noticed a weak  signature of absorption feature around 50 keV in the residuals of the  
AstroSat  spectrum. And, since the source appears  to be marginally detectable above background at around 
50 keV with $1-2\sigma$ detection limit for both the AstroSat observations.  We, therefore, attempted to include
an additional absorption line around 49 keV in the combined spectral model for both the continuum. 
We could not detect  presence of any significant absorption feature in the  spectrum around 49--50 keV in the
AstroSat data as overall difference in $\chi^2$, $\Delta\chi^2\approx 5$  was found. 
As a result, we have not included this feature in further analysis.

%
%
%
\begin{deluxetable}{cccccc}
\tablecolumns{6}
\tablecaption{Spectral parameters obtained for phase averaged spectrum obtained from two AstroSat observations} 
\tablehead{\colhead{Parameter}& \colhead{Units}& \colhead{Obs-1(FD-cutoff)}& \colhead{Obs-2(FD-cutoff)} & \colhead{Obs-1(compTT)}& \colhead{Obs-2(compTT)}}
\startdata
\hline
\textbf{Parameter values with 16 keV line} \\
\hline
Phabs(nH) & 10$^{22}$ cm$^{-2}$ & $0.87^{+0.07}_{-0.07}$  & $0.83^{+0.07}_{-0.07}$  & $0.77^{+0.06}_{-0.06}$ & $0.62^{+0.05}_{-0.05}$ \\
bbody(kT) & keV & $0.62^{+0.04}_{-0.05}$ & $0.69^{+0.04}_{-0.05}$ &- &- \\
bbody(Norm) & ph keV$^{-1}$ cm$^{-2}$ s$^{-1}$ & $1.98^{+0.59}_{-0.51}\times 10^{-3}$ & $8.02^{+2.95}_{-3.04}\times 10^{-4}$ &- & - \\
\hline
gauss(Line E) & keV & 6.5 (fixed) & 6.5 (fixed)  & 6.5 (fixed) & 6.5 (fixed) \\ 
gauss(Sigma) & keV & 0.30 (fixed) & 0.30 (fixed)  & 0.30 (fixed) & 0.30 (fixed)\\
gauss(Norm)  & ph~cm$^{-2}~s^{-1}$ & $5.74^{+6.07}_{*}\times 10^{-4}$  & $3.94^{+3.03}_{-2.95}\times 10^{-4}$  & $5.69^{+6.88}_{*}\times 10^{-4}$  & $3.16^{+2.69}_{-2.52}\times 10^{-4}$ \\
\hline
gauss(Line E) & keV & $16.33^{+0.49}_{-0.63}$ & $16.12^{+1.36}_{-1.25}$ &$15.00^{+1.56}_{-1.84}$ & $15.56^{+0.98}_{-1.72}$ \\ 
gauss(Sigma) & keV &  2.34 (fixed)  &  $2.22^{+1.37}_{-0.71}$ & $3.43^{+1.50}_{-1.08}$  &  $2.34^{+1.71}_{-1.03}$  \\
gauss(Norm)  & ph~cm$^{-2}~s^{-1}$ & $2.19^{+0.49}_{-0.51}\times 10^{-3}$  & $4.42^{+2.99}_{-3.10}\times 10^{-4}$  & $4.05^{+6.32}_{-1.84}\times 10^{-3}$  & $6.60^{+13.12}_{-3.48}\times 10^{-4}$ \\
\hline
power-law(PI) &- & $0.83^{+0.17}_{-0.23}$ & $0.98^{+0.13}_{-0.12}$ & - & - \\
$E_{cut}$ & keV & $12.79^{+4.25}_{-5.21}$ & 12.79 (fixed) & - &  - \\
$E_{fold}$ & keV & $7.78^{+0.41}_{-0.37}$ & $11.52^{+1.36}_{-1.08}$ & - & -  \\
power-law(Norm) at 1~keV & ph keV$^{-1}$ cm$^{-2}$ s$^{-1}$ & $5.74^{+1.38}_{-1.31} \times 10^{-2}$ & $3.98^{+1.26}_{-0.95}\times 10^{-2}$ & - & - \\
\hline
compTT(T0)  & keV & - & - & $0.42^{+0.04}_{-0.03}$ & $0.49^{+0.03}_{-0.03}$ \\
compTT(kT)  & keV & - &-  & $5.17^{+0.25}_{-0.17}$ &$6.92^{+0.62}_{-0.49}$  \\
compTT(tau) & - & - &- & $6.05^{+0.26}_{-0.53}$ & $5.21^{+0.22}_{-0.30}$ \\
compTT(Norm)  & -  & - &- & $4.78^{+0.29}_{-0.43}\times 10^{-2}$ & $1.72^{+0.12}_{-0.12}\times 10^{-2}$ \\
\hline
cyclabs(Line E0) & keV & $30.48^{+0.33}_{-0.34}$ & $30.68^{+0.45}_{-0.44}$  & $30.46^{+0.32}_{-0.28}$ & $30.30^{+0.40}_{-0.37}$ \\ 
cyclabs(Width) & keV & $3.25^{+1.14}_{-0.93}$  &  $5.89^{+1.38}_{-1.26}$ & $3.33^{+1.10}_{-1.01}$  &  $7.12^{+1.33}_{-1.22}$ \\
cyclabs(Opt. Depth)  & -   & $1.03^{+0.14}_{-0.11}$  & $1.42^{+0.18}_{-0.18}$  & $1.02^{+0.12}_{-0.12}$  & $1.52^{+0.20}_{-0.18}$ \\
\hline
Flux(3--80 keV) & erg cm$^{-2}$ s$^{-1}$  & $1.62\times 10^{-9}$ & $8.08\times 10^{-10}$  & $1.61\times 10^{-9}$ & $8.07\times 10^{-10}$\\
\hline
$\chi^2$   & -         & 429.09 & 456.13 & 426.53 & 463.03 \\
$\chi^2_{{red}}$ (dof) &-  & 0.962(446) & 1.016(449) & 0.956(446)  & 1.031(449)  \\
\hline
\hline
\textbf{Parameter values without 16 keV line} \\
\hline
Phabs(nH) & 10$^{22}$ cm$^{-2}$ & $0.87^{+0.04}_{-0.04}$  & $0.76^{+0.04}_{-0.04}$  & $0.77^{+0.06}_{-0.06}$ & $0.62^{+0.05}_{-0.05}$ \\
bbody(kT) & keV & $0.65^{+0.03}_{-0.04}$ & $0.73^{+0.03}_{-0.03}$ &- &- \\
bbody(Norm) & ph keV$^{-1}$ cm$^{-2}$ s$^{-1}$ & $2.04^{+0.33}_{-0.32}\times 10^{-3}$ & $1.17^{+0.19}_{-0.19}\times 10^{-3}$ &- & - \\
\hline
gauss(Line E) & keV & 6.5 (fixed) & 6.5 (fixed)  & 6.5 (fixed) & 6.5 (fixed) \\ 
gauss(Sigma) & keV & 0.30 (fixed) & 0.30 (fixed)  & 0.30 (fixed) & 0.30 (fixed)\\
gauss(Norm)  & ph~cm$^{-2}~s^{-1}$ & $9.20^{+5.86}_{-5.74}\times 10^{-4}$  & $6.56^{+2.89}_{-2.83}\times 10^{-4}$  & $1.29^{+5.07}_{*}\times 10^{-4}$  & $2.58^{+2.50}_{-2.43}\times 10^{-4}$ \\
\hline
Power-law(PI) &- & $0.93^{+0.05}_{-0.05}$ & $0.88^{+0.06}_{-0.06}$ & - & - \\
$E_{cut}$  & keV & $20.88^{+3.46}_{-2.27}$ & 20.88 (fixed) & - &  - \\
$E_{fold}$ & keV & $7.40^{+0.46}_{-0.39}$ & $10.65^{+1.31}_{-1.01}$ & - & -  \\
power-law(Norm) at 1~keV & ph keV$^{-1}$ cm$^{-2}$ s$^{-1}$ & $5.34^{+0.64}_{-0.62} \times 10^{-2}$ & $2.52^{+0.39}_{-0.35}\times 10^{-2}$ & - & - \\
\hline
compTT(T0)  & keV & - & - & $0.40^{+0.03}_{-0.03}$ & $0.48^{+0.02}_{-0.03}$ \\
compTT(kT)  & keV & - &-  & $5.58^{+0.27}_{-0.22}$ &$7.24^{+0.63}_{-0.49}$  \\
compTT(tau) & - & - &- & $6.30^{+0.17}_{-0.17}$ & $5.35^{+0.18}_{-0.19}$ \\
compTT(Norm)  & -  & - &- & $4.55^{+0.25}_{-0.23}\times 10^{-2}$ & $1.67^{+0.11}_{-0.11}\times 10^{-2}$ \\
\hline
cyclabs(Line E0) & keV & $29.57^{+0.35}_{-0.35}$ & $29.99^{+0.44}_{-0.43}$  & $30.23^{+0.26}_{-0.26}$ & $30.16^{+0.36}_{-0.34}$ \\ 
cyclabs(Width) & keV & $6.93^{+0.96}_{-0.86}$  &  $8.09^{+1.10}_{-0.97}$ & $6.54^{+1.11}_{-0.96}$  &  $8.14^{+1.11}_{-1.01}$ \\
cyclabs(Opt. Depth) & -   & $1.17^{+0.12}_{-0.11}$  & $1.62^{+0.18}_{-0.16}$  & $1.26^{+0.12}_{-0.10}$  & $1.70^{+0.19}_{-0.17}$ \\
\hline
$\chi^2$   & -         & 465.21 & 473.50 & 484.94 & 463.03 \\
$\chi^2_{{red}}$ (dof) &-  & 1.036(449) & 1.048(452) & 1.080(449)  & 1.069(452)  \\
\hline
\enddata
\tablecomments{Errors quoted for each parameters above are with 90\% confidence range. *--parameter pegged at hard limit 0}  
\end{deluxetable}
%
\begin{figure}
\centerline{\includegraphics[scale=0.36,angle=-90]{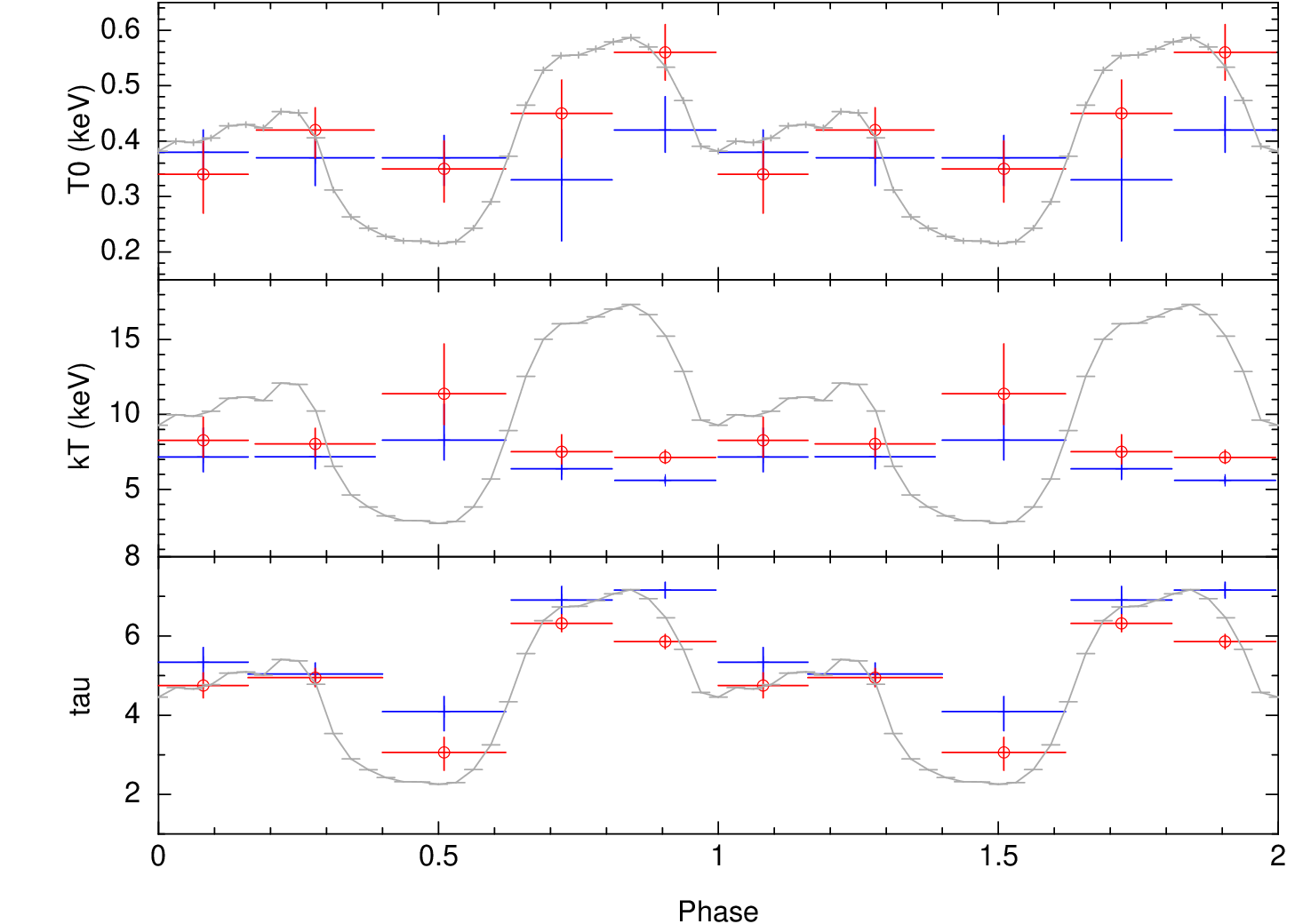}
\quad\includegraphics[scale=0.36,angle=-90]{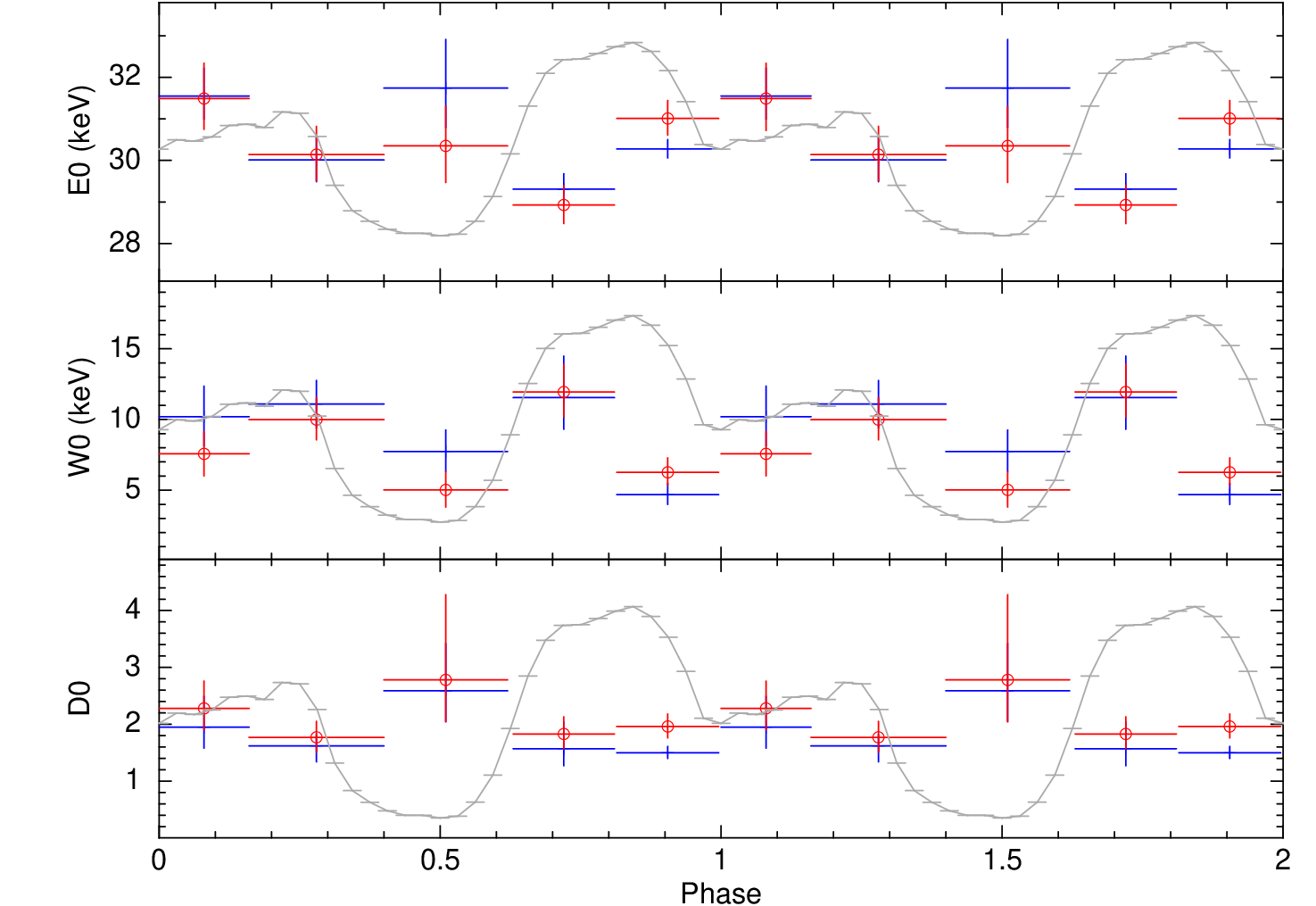}}
\caption{Spectral parameters with 90\% errors as derived from phase resolved spectroscopy of  AstroSat observation-1 
(Blue colored points)  \& observation-2 (Red colored points) are shown for the continuum defined
by thermal  Comptonization compTT-model (left panel) and the 
cyclotron absorption line  defined by cyclabs-model (right panel) with respect to different pulse phases as 
defined in figure-3 and can be seen from the overlay-pulse of observation-2.}   
\label{phs2}
\end{figure}
%

Studies of spin-pulse phase resolved spectroscopy were performed from two AstroSat 
observations of 2018 outburst during its declining phase with a source intensity changed by 
a factor of $\sim2$.  The  spectral data from SXT and LAXPC20  from two AstroSat observations, 
were extracted corresponding to the 5-different  pulse-phase intervals considered 
as marked in Figure~\ref{pulsea}.  
The phase intervals were selected primarily  for easy comparison of AstroSat results with the 
results of 2014 outburst observed by NuSTAR \citep{bhargava19}, and
to capture variations in the structures of  secondary and primary pulses of  the pulse profile 
in  reference to the soft X-rays  energy band  as marked in Figure~\ref{pulsea} (left-panel) 
and to follow variations in the higher energy bands in the same 5 phase-intervals.  
The  primary and the secondary pulse were divided in 2 phase-intervals each, 
while  a single interval was chosen to include the valley region  (Ph3), in between the 
secondary-pulse and the primary-pulse (Figure~\ref{pulsea}). The valley region so defined is found to become 
gradually  wider and flatter at higher energies than at lower X-ray energy bands as can be seen in Figure~\ref{pulseb}. 
Therefore, the second phase interval at higher energies  appears to capture larger extent of the narrower 
secondary pulse at higher energies than at lower energies (Figure~\ref{pulsea}, Figure~\ref{pulseb}).

The compTT-model was preferred over the FD-cutoff model for 
phase resolved spectroscopy, as parameters of cutoff model such as  E-cut has no clear physical meaning, 
although it might be related to the strength of the magnetic field \citep{makishima90}, where as E-fold has been 
associated by some authors \citep{unger92} with the plasma temperature in the emission region. The thermal 
Comptonization compTT-model on the other hand, offers  significant  physical details about plasma distributions 
in  the pulsar accretion column in terms of its plasma temperature, its optical depth in addition to its input 
photon Wien temperature  over the different pulse-phases. 
We, thus, studied phase resolved spectroscopy for Cepheus X-4  for the first 
time employing a compTT-model without a black-body component for the spectral continuum. 
The black-body temperature variation, therefore,  was not 
considered here, like it was presented  earlier along with FD-cutoff model by \citet{bhargava19}.  
All the important 
spectral features of the Cepheus X-4, as observed in the phase-averaged spectrum of  AstroSat were 
also  present  in the  phase-resolved spectra of the pulsar.  As regards to the Fe-line, its  
energy and width were frozen to the values as in the phase-averaged case for all the 5-phases considered, 
as extracted data for the 5 individual phase intervals could not constrain the Fe-line 
parameters  during the spectral fit.  Thus, all these  spectra
were well fitted  using absorbed compTT model for the continuum  along with a  
fixed Fe-emission line and a cyclotron absorption feature defined by cyclabs-model.  
We have not included  16 keV feature  for AstroSat phase resolved spectroscopic studies,
to avoid any interference of this feature with cyclotron-line parameters. 
For phase resolved spectrum, the significance of the feature is also low because of division of data in 5-phases resulting 
in lower signal to noise ratio.
Our main motivation is to study the fundamental line through phase-resolved 
spectroscopy. Therefore, results obtained from  analysis of pulse phase-resolved spectra corresponding to 5 different 
pulse-phases of the pulsar are presented in Figure~\ref{phs2}.  Spectral parameters derived for the continuum 
fitted  with a compTT-model are plotted in left-panel and  cyclotron line parameters  are plotted on 
the right-panel of the Figure~\ref{phs2}. These results from both the observations  were plotted 
on the same graph along with extent of each interval along with overlay spin-pulse of observation-2 for direct 
comparison of phase-resolved spectral parameters and studying relative variability at two different source intensity.
  
It is noticed that input soft photon Wien temperature T0 varies, in the these 5 pulse phases, covering the range 
including its 90\% error limits from 0.22--0.48 keV and 0.27--0.61 keV  for two observations respectively. 
It has a maximum at the falling edge of the primary-pulse around  
phase 0.9 and minimum near the rising edge of the  secondary pulse around phase 0.1  
for the both AstroSat observations as seen in Figure~\ref{phs2}. 
We observed plasma temperature kT  though marginally higher for observation-2 than for observation-1, but
remained consistently stable in all phases within its 90\% error limits. It covers 
a range 5.41--10.66 keV for the observation-1 and 6.74--14.71 keV for observation-2 within its error limits. 
The plasma optical depth $\tau$ varied in the range 3.61--7.36  for observation-1 and 2.61--6.54  for observation-2 
with  relatively lower range of variation and following  pulse 
amplitude variation with phase with minimum near pulse-minima at phase 0.5 and  maximum 
corresponding to primary-pulse-maxima near the phase 0.7 as can be seen in Figure~\ref{phs2} (left panel).
The cyclotron line  energy  within 90\% error limits, was found to vary in the range 
from  28.93--32.91 keV and  28.48--32.34 keV for two observations respectively. The  minimum is
near the rising edge of the primary around phase 0.7 and maximum is  around rising edge of the secondary
around phase 0.1.  
The centroid energy for the cyclotron line for both  AstroSat observations,  showed 
variation with the pulse-phase as seen in Figure~\ref{phs2} (right panel). 
The cyclotron line-widths showed overall indication of variations with pulse amplitude in phase but
remained consistent for both the observations within 90\% error limits.  The line-widths was found 
to vary in the range 3.99--12.77 keV and 3.81--13.91 keV over the phase for the two observations 
respectively with some indication of variation.  The  optical-depths of the cyclotron line were 
found in the range 1.40--2.49  and 1.51--2.76  over the complete phase.  
The optical-depth, however, remains stable  overall with phase except for some indication of  increase near phase 0.5 
for these two  AstroSat observations.\\

%
\begin{deluxetable}{ccccc}
\tablecolumns{5}
\tablecaption{Pulse period measurements of Cepheus X-4 }
\tablehead{\colhead{Observation} &\colhead{Telescope} &\colhead{MJD} &\colhead{Pulse-Period${^a}$} &\colhead{Ref$^{c}$}} 
\startdata
1988 April & GINGA$^{b}$ & 47263.5  & $66.2490 \pm 0.0001$  & [1]\\\\
1993 June  & ROSAT & 49160.0  & $66.2552 \pm 0.0007$  & [2]\\\\
1993 June  & BATSE & 49163.0  & $66.2510 \pm 0.0009$  & [3]\\
           &       & 49167.0  & $66.2497 \pm 0.0009$  &   \\
           &       & 49169.0  & $66.2511 \pm 0.0009$  &   \\
           &       & 49171.0  & $66.2502 \pm 0.0008$  &   \\
           &       & 49173.0  & $66.2501 \pm 0.0007$  &   \\\\
1997 July  & BATSE & 50633.0  & $66.2713 \pm 0.0009$  & [3]\\
           &       & 50635.0  & $66.2733 \pm 0.0009$  &    \\
           &       & 50637.0  & $66.2749 \pm 0.0009$  &    \\
           &       & 50639.0  & $66.2743 \pm 0.0009$  &    \\
           &       & 50641.0  & $66.2769 \pm 0.0009$  &    \\
           &       & 50643.0  & $66.2775 \pm 0.0009$  &    \\\\
1997 July  & RXTE  & 50647.0  & $66.278  \pm 0.017 $  & [3] \\
           &       & 50654.0  & $66.273  \pm 0.017 $  &     \\\\
2002 July  & RXTE  & 52450.7  & $ 66.30  \pm 0.01$    & [4] \\
           &       & 52451.3  & $66.30   \pm 0.01$    &     \\ 
           &       & 52451.7  & $66.296  \pm 0.004$   &     \\\\
2014 June  & NuSTAR & 56827.38 & $66.3352 \pm 0.0003$ & [5] \\
2014 July  &        & 56839.87 & $66.3336 \pm 0.0002$ &      \\\\
2018 July  & AstroSat & 58301.61850 & $66.35080 \pm 0.00008$ & [6] \\
2018 July  &          & 58307.40211 & $66.35290 \pm 0.00011$ &     \\
\enddata
\tablecomments{${^a}$Errors with 68\% confidence range for each parameter. 
${^b}$ Helio-centric corrected time.
${^c}$References [1] Koyama et al. (1991) [2] Schulz et al. (1995) 
[3]  Wilson et al. (1999) [4] McBride et al.(2007) [5] Vybornov et al.(2017) [6] Present Work}
\end{deluxetable}
%
%
\section{Discussion}
\subsection{Long term Spin-down of the Pulsar }
The measurements of spin-up and spin-down rate in X-ray pulsars offers scope for in-depth 
studies of underlying mechanism for transfer of angular momentum to the neutron star by 
accreting matter \citep{bildsten97}.  It  can  serve as a tool to investigate the 
interaction of accreting plasma with magnetosphere of rotating neutron stars and the
nature of the different  torques acting  on the neutron star \citep{nelson00, dai16}. 
In the case of  a wind-fed accretion in a high mass X-ray binary, if the specific angular 
momentum of gravitationally captured matter is sufficiently high, then an accretion disk can 
possibly be formed around the neutron star, as seen in the cases of  A0535+262 \citep{finger96, finger98}, 
EXO 2030+375 \citep{angelini89}, 4U 0115+63 \citep{soong89, heindl99, dugair13}, V0032+53 \citep{qu05}, 
4U 1907+09 \citep{mukerjee01}, and GRO J2058+42 \citep{mukerjee20}.  Accretion, otherwise  can  occur 
through different regimes of a quasi-spherical accretion on to  the neutron star \citep{burnard83, shakura12}.
Measurements of pulse period variations in such systems can provide scope for understanding 
different processes responsible for the angular momentum transfer, the structure of the accretion 
flows in binary systems, their dependence on system parameters, X-ray luminosity, etc. 
Different physical mechanisms for torques acting on X-ray pulsars  have been proposed 
to explain these variations \citep{ghosh79, love95, rappa04}. 

\begin{figure}
\centerline{\includegraphics[scale=0.45,angle=-90]{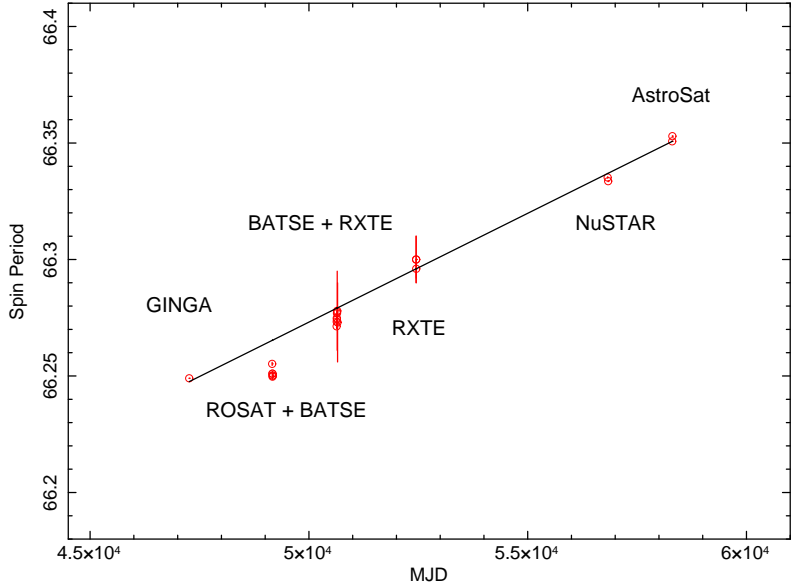}}
\caption{Pulse period  of Cepheus X-4 during different outburst observed by 
 different observatories over the last 30 years.}
\label{pdot}
\end{figure}

For studies of variation of spin-period  of Cepheus X-4, we compiled all the period measurements 
of the Cepheus X-4 observed during various outbursts over the last 30 years by different observatories 
(Table-2).  All these data are plotted with a linear fit and shown in Figure~\ref{pdot}. 
Two types of variations in the spin-period are seen in the Cepheus X-4 during all these
observations as can be seen from  Figure~\ref{pdot}.  Firstly, the source showed luminosity dependent 
spin-period variations during each  outburst which gives rise to the spread of data points in Figure~10 
about the mean trend, as expected from accretion 
torque theory \citep[][and references therein]{ghosh79, bildsten97, malac20}.  
Whereby during an outburst, depending on change in the mass 
accretion rate and hence due to  proportionate change in transfer of angular momentum,  it affects change in 
the pulsar spin period and its rate  as seen in Cepheus X-4. Similar changes were also observed for  many other 
Be-binaries during an outburst \citep{sugizaki17, finger96, parmar89b}.
During declining phase of  2018 outburst, the observed rate of change in spin period of Cepheus X-4 are 
relatively higher at higher source luminosity  $\dot\nu=(-2.1\pm0.8)\times10^{-12}$ Hz s$^{-1}$
for the first AstroSat observation compared  to the second observation  
$\dot\nu=(-1.6\pm0.8)\times10^{-12}$ Hz s$^{-1}$ at lower source luminosity 
as expected from the accretion torque theory.  Additionally,  a long term secular decrease in the spin period 
of the pulsar was also observed between each of these outbursts as shown by the straight line fit to all the data.  
The  straight-line fit to the data showed a continuous decrease  in the pulsar spin-period at a rate of 
${1.081\pm 0.002} \times 10^{-10}~s~s^{-1}$ ($\dot\nu = -(2.455\pm 0.004)\times10^{-14}$ Hz s$^{-1}$) 
over the long duration of about 30 years, which we investigate in detail below. For Ginga observation  only  
helio-centric correction was applied to the measured value of the spin-period. 
This average spin-down is consistent with the earlier  reported measurement of BATSE 
$\dot{P} = 1.82\times 10^{-10}$~s~s$^{-1}$  ($\dot\nu = -4.14 \times10^{-14}$ Hz s$^{-1}$) 
between two outbursts of 1993 and 1997, \citep{wilson99}.

The  observed continuous slowing down behavior of pulsars could be due to
resulting centrifugal inhibition of accretion, which occurs at the end of the outburst. 
During the declining phase of an outburst, there is a gradual decrease of accretion on to the neutron star 
leading to the reduction of source luminosity.  As the accretion rate decreases, the ram pressure decreases, 
the magnetosphere expands  and  magnetospheric radius $r_{m}$ extends closer to the co-rotation radius 
$r_{co}$ where the angular velocity of Keplerian motion is equal to that of the neutron star. 
As the magnetosphere grows beyond the co-rotation radius, centrifugal force prevents material
from entering the magnetosphere, and thus accretion onto magnetic poles ceases 
\citep{pringle72, lamb73, illar75}. Consequently, no X-ray pulsation
is expected. This is commonly known as the propeller effect, because accreting matter is 
likely to be ejected in the presence of a strong magnetic field.
The in-falling material is flung away, carrying part of the angular momentum extracted from the neutron
star, resulting in slowing down of the neutron star. 
As this mechanism occurs at  relatively much lower  source luminosity at about 
$L_{X}\mathrm{(min)}= 1.3 \times 10^{34}$~erg~s$^{-1}$ as shown below for Cepheus X-4 and hence at much 
lower mass accretion rate at the end of the  source outburst.  Therefore, change  of angular momentum 
of the neutron star due to onset of propeller effect is at a much lower rate 
of spin-down
($\dot\nu = -(2.455\pm 0.004) \times 10^{-14}$~Hz~s$^{-1}$), which is two orders of magnitude lower 
than that during  the outburst as derived from all the  available observations of Cepheus X-4. 
Now, prior to detailed investigations on the Propeller effect, we  derive some of the 
required parameters associated with the neutron star as below. 

In the presence of a strong magnetic field of a neutron star of an accreting X-ray binary, 
the plasma electrons in the accretion column, get quantized in the Landau states.  X-ray photons  
get scattered from these quantized plasma electrons in the strong magnetic field 
close to the surface of an accreting X-ray pulsar and  produce  cyclotron resonant scattering 
features. It results in appearance of a complex absorption features at a particular energy 
in the X-ray spectrum depending on strength of  magnetic field of the neutron star 
\citep{schw17}.  Alternatively,  reflection of X-rays from the atmosphere of the neutron star can also
produce  cyclotron absorption features \citep{pout13}. The complex shape of the  
cyclotron absorption feature  depends on  many factors. These include physical condition of the 
line-forming region, strength of the local magnetic field, nature of the accretion column,  
source luminosity and  pulse-phase of an accreting X-ray pulsar \citep{nishi03, nishi05, nishi13}. 
From the measurement of centroid energy of the cyclotron absorption feature, 
the strength of the surface magnetic field $B$ of a neutron star  can be 
obtained using the following equation
\begin{equation}
	E_{0} \simeq  11.57  n  \left(\frac{1}{1+z} \right)   \left(\frac{B}{10^{12}G} \right)\;  \mbox{keV}.
	\label{ecycl}
\end{equation}
We assume that the observed  cyclotron absorption  feature at $E_{0}$ is associated with its fundamental 
line ($n = 1$), and considering a gravitational red-shift of $z = 0.3$ at the surface of a 
typical neutron star having  a mass of $1.4 M_{\odot}$ and a radius of 10 km. 
From the measurements of the centroid energy of $30.46^{+0.32}_{-0.28}$ keV for observation-1, 
and $30.30^{+0.36}_{-0.34}$ keV for observation-2, derived from the phase averaged spectrum using compTT model (Table~1), 
we could  thus establish the strength  of the magnetic field of the pulsar as 
$(3.42^{+0.04}_{-0.03}) \times 10^{12}$ G and  $(3.40^{+0.04}_{-0.04}) \times 10^{12}$ G, respectively.
The cyclotron line energies  agree within their error limits, shows consistency during 
two observations even when the source luminosity changed  by a factor of $\sim~2$.

If all the potential energy of the accreted mass released during the  
outburst is converted into radiation then  the  luminosity  can be expressed as,
\begin{equation}
 L_{37} = 0.133 \times {\dot{M}_{16}} \left(\frac{M}{M_{\odot}}\right){R_{6}}^{-1} \quad \mbox{erg s}^{-1},
 \label{l37}
\end{equation}
where $L_{37}$ is the luminosity in units of $10^{37}$ erg s$^{-1}$ and $R_6$ is the
radius in units of $10^6$ cm, $\dot{M}_{16}$ is the mass accretion rate in
$10^{16}$ g s$^{-1}$ and M is the mass of neutron star.  The mass accretion 
rate $\dot{M}$ can be calculated from the 
observed luminosity, which is derived by measurements of flux from  spectral model 
for AstroSat spectrum (Table~1), and known distance of the star. 
The Cepheus X-4 luminosity during the AstroSat observations were 
estimated to be $2.04 \times 10^{37}$~erg~s$^{-1}$  
and  $1.02 \times 10^{37}$~erg~s$^{-1}$ respectively,
in 3--80 keV energy band. This was derived using distance measurements of Gaia
as $10.2^{+2.2}_{-1.6}$ kpc \citep{malac20} and  estimated flux from 
AstroSat observations (Table~1).  
The estimated value of $\dot{M}$ during the  AstroSat observations were 
$11.0 \times 10^{16}$ g s$^{-1}$ and $5.5 \times 10^{16}$ g s$^{-1}$ respectively.
It is therefore inferred that
the accretion process was intensively active during AstroSat observations.

The magnetospheric radius $r_{m}$, of the neutron star can be  expressed 
in terms of its mass accretion rate and strength of magnetic field of neutron star 
by the Equation-6.18, page 158 of \citet{frank02}
\begin{equation}
 r_{m} \simeq  5.2~\mu_{30}^{4/7} {\dot{M}_{16}}^{-2/7} {m}^{-1/7}
 \times 10^{8} \, \mbox{cm} = 5.0 \times 10^{8} \, \mbox{cm},
	\label{eq:rmag}
\end{equation} 
where $m= M/{M_{\odot}} = 1.4$ is the mass ratio of neutron star, 
$\dot{M}_{16} = 11.0$ is its mass accretion rate expressed in units of $10^{16}$ g s$^{-1}$  
and $\mu_{30}$ is magnetic moment of the neutron star expressed in units of $10^{30}$ G cm$^{3}$.  
The magnetic moment ($\mu)$ of the neutron star for a dipole-like field configuration 
is given by ${{\mu} = B  R^3}$, where $B$ is surface polar magnetic field strength and 
$R$ is the radius of the neutron star. From the AstroSat measurements of cyclotron 
resonance scattering  feature, we have derived ${B = 3.4 \times 10^{12}}$ G, which 
gives ${\mu}_{30} =  3.4$ for a 10 km radius of the neutron star.
Thus, all these parameters are  derived from observed values of neutron star magnetic 
field and source luminosity respectively from AstroSat observation.

%

The radius,  where the spin angular velocity of neutron star is equal to the Keplerian 
angular velocity of matter, is defined as the co-rotation radius $r_{co}$ of an X-ray pulsar
and can be expressed as
\begin{equation}
	r_{co} = 1.7 \times 10^{8} P^{2/3} \left(\frac{M}{1.4 M_\odot} \right)^{1/3} \, \mbox{cm}
       = 2.8 \times 10^{9} \, \mbox{cm}.
	\label{eq:rco}
\end{equation} 
Using measured value of the spin period of the neutron star from AstroSat observations 
$P=66.35$ s and assuming a neutron star mass of $M = 1.4 M_{\odot}$, 
one  can obtain the co-rotation radius $r_{co}$ for Cepheus X-4 as 
$2.8 \times 10^{9}$ cm (Equation-\ref{eq:rco}). 
Now, by applying the condition $r_{m} = r_{co}$, that is 
equating Equation-\ref{eq:rmag} and Equation-\ref{eq:rco} and utilizing Equation-\ref{l37}, one can derive 
the minimum luminosity $L_{X}$ (min) at which 
an X-ray pulsar turns off due to the centrifugal inhibition of accretion, so called  the
propeller effect, is given by \citep{illar75}
\begin{equation}
	L_{X}\mathrm{(min)} \approx 2 \times 10^{37} {\left(\frac{R}{10^6~cm} \right)}^{-1} \left(\frac{M}{1.4 M_\odot} \right)^{-2/3} \mu_{30}^{2} \left(\frac{P}{1 s} \right)^{-7/3}~~\hbox{erg~s}^{-1},
\end{equation} 
where R, M, ${\mu}$ and ${P}$ are radius, mass, magnetic moment and
spin period of the neutron star respectively. 
Using derived values of $\mu_{30}$, ${P}$  and ${M = 1.4 M_{\odot}}$, 
one can obtain $L_{X}\mathrm{(min)}= 1.3 \times 10^{34}$~erg~s$^{-1}$ for
Cepheus X-4. The pulsar luminosity must go down below
this limiting value for the propeller effect to prevail. 
As a limiting case, corresponding to the minimum luminosity  of 
$L_{X}\mathrm{(min)} = 1.3 \times 10^{34}$~erg~s$^{-1}$ below which the 
pulsar turns off due to propeller effect as derived above, the corresponding $\dot{M}$ 
can be estimated as ${\dot{M} = 7.0 \times 10^{13} ~g~s^{-1}}$ using Equation-4. Similarly, in the
propeller regime where $r_{m} = r_{co}$ (where $r_{co}$ can be estimated using Equation-6), 
one can estimate the corresponding value of  magnetic moment 
$\mu$ of the neutron star  as $\mu_{30} = 1.7$ (using Equation-5), 
which corresponds to  magnetic field  of $B = 1.7 \times 10^{12}$~G for a 10 km radius
neutron star.

Many  researchers have investigated this phenomenon of propeller effect over 
the period \citep{pringle72, illar75, davies81, ikhsanov01, shakura12, malac20}. 
It was suggested that the propeller phase of the pulsar can be divided 
into two sub-phases (a) a supersonic propeller phase or (b) a subsonic propeller 
phase \citep{davies81}. This classification was made
depending on the angular velocity of the pulsar ($\Omega_{s}$). For the case of  
supersonic propeller phase  ${r_{c}\Omega_{s} \gg c_{s}(r_{c})}$, 
where $r_{c}$ is the inner boundary of the accretion disk and $c_{s}(r_{c})$ 
is the sound speed at the radius $r_{c}$.  
Let us therefore consider these two cases of propeller
regime separately for Cepheus X-4 and infer the most likely scenario  based on observed value of $\dot{P}$.

(a) Supersonic Propeller Phase: The spin down rate during the supersonic propeller phase
can be obtained  using the  equation  derived  by \citet{ikhsanov04} and expressed as
\begin{equation}
 \tau_{c}  = \frac{P}{2\dot{P}} \simeq  5.12 \times 10^{13}~~\mu_{30}^{-3/7}~~{{I}_{45}}~~\dot{M}_{16}^{-11/14}~~V_{7}^{1/2} \, 
     \mbox{~s}. 
	\label{eq:pppdot}
\end{equation} 
Here $V_{7}$ is  the relative velocity of the star and surrounding medium in the units of $10^{7}$~cm~s$^{-1}$. Using
$V_{7} =~1$, $\mu_{30}=1.7$ G~cm$^{3}$ and  ${\dot{M} = 7.0 \times 10^{13}}$ g~s$^{-1}$  we
get $\dot{P} = 1.6 \times 10^{-14}$~s~s$^{-1}$ ($\dot\nu = -3.6 \times10^{-18}$ Hz s$^{-1}$) 
for the supersonic  propeller case.

(b) Subsonic Propeller Phase: The spin down rate during the subsonic propeller phase
is as derived by \citet{ikhsanov04}
\begin{equation}
 \dot{P} = \left(\frac{P^{3}L_\mathrm{ssp}}{4\pi^2I}\right)  \simeq  2.5 \times 10^{-11} \mu_{30}^{2} {m}^{-1} {{I}_{45}}^{-1} 
  \, \mbox{~s~s$^{-1}$} = 0.52 \times 10^{-10} \, \mbox{~s~s$^{-1}$},
	\label{eq:dp}
\end{equation} 
where $m= M/{M_{\odot}} = 1.4$ is the mass ratio of neutron star, $I_{45}$ is moment of inertia of 
the neutron star expressed in the units of $10^{45}$ g cm$^2$ and  $L_\mathrm{ssp}$ is the 
luminosity  of the pulsar in the subsonic propeller regime expressed by
\citep{davies81, ikhsanov04}. 
\begin{equation}
L_{ssp} =  1\times 10^{36} \mu_{30}^{2} {m}^{-1} P^{-3} \mbox{~erg~s}^{-1} .
	\label{eq:lsup}
\end{equation} 
Using these above parameters established for Cepheus X-4 in the propeller regime, 
the corresponding  value of $\dot{P}$ was established and found to be
${\dot{P}= 0.52 \times 10^{-10}~s~s^{-1}}$ (${\dot\nu = -1.18 \times10^{-14}}$ Hz s$^{-1}$). 
This value is comparable to the observationally derived 
$\dot{P} = (1.081\pm0.002) \times 10^{-10}$~s~s$^{-1}$ ~($\dot\nu = -(2.455\pm 0.004)\times10^{-14}$~Hz~s$^{-1}$)
within underlying parameter uncertainty.

We also cross-verified the subsonic case described by \citet{malac20}
and derived the $\dot{P}$ as described by their  Equation-12
\begin{equation}
 \dot{P}  \simeq  1.16 \times 10^{-11}~~{(2\pi)}^{-1}~~\Pi_{sd}~~\mu_{30}^{13/11}~~\dot{M}_{16}^{3/11}~~P \, 
     \mbox{~s~s$^{-1}$} = 0.59 \times 10^{-10} \times \Pi_{sd} \, \, \mbox{~s~s}^{-1},
	\label{eq:ppdot}
\end{equation} 
Both of these estimates of $\dot{P}$ showed consistency with the value derived value from observations favoring  the
case of subsonic propeller regime of the pulsar. The comparison of observed  value with Equation-\ref{eq:ppdot}, suggest that
the value of  dimensionless parameter would be $\Pi_{sd}$ = 1.8 for the case of Cepheus X-4.  
The parameter  ${\Pi_{sd}}$  is a combination of dimensionless parameters \citep{shakura14} and its
derived values from observations  were found  to be between 5--10 for some of the other X-ray pulsars studied \citep{postnov15}.
This result therefore suggest that the pulsar enters subsonic propeller regime after each outburst activity
observed with 4--5 years interval and hence showed continuous slowing down of its spin period. Thus, 
using this established  spin-down rate one can predict  pulse-period of the pulsar for an epoch when 
the next outbursts would occur.   

\subsection{Pulse Phase-averaged Spectrum}

The cyclotron line energy was measured  for Cepheus X-4 by various observatories
during different outbursts  observed  over the long period of more than 30 years since its discovery 
in 1988 by Ginga. It was first measured by Ginga and then followed by RXTE, 
NuSTAR and most recently by  AstroSat.  
The  measurements of  cyclotron line parameters from phase-averaged spectrum obtained during different 
outburst of the source  are compiled in Table~3 along with  observation epoch and the observed source flux.  
The Ginga spectral data were fitted by a combination of an absorbed power-law along with cyclotron  
resonance scattering  (cyclabs) as its cutoff model \citep{mihara91} and measured  centroid line energy. 
While for  rest of the measurements,  
spectral data were modeled by absorbed Fermi-Dirac cutoff model \citep{tanaka86} combined  with the same cyclotron scattering 
(cyclabs) model. The use of different continuum  and absorption line models affect measurements  of 
cyclotron line parameters  marginally.  
Therefore, for the purpose of comparison we report here the values obtained in previous studies,  derived
using the same model except for the Ginga, as tabulated in Table~3 along with respective references. 
The  results from  simultaneous Suzaku observation of 2014 outburst 
\citep{jaiswal15} with NuSTAR was not considered here for comparison, since cyclotron feature in the spectrum 
was modeled employing  gabs-model  although its continuum was modeled using same Fermi-Dirac cutoff model.

%
\begin{deluxetable}{cccccccc}
\tablecolumns{8}
\tablecaption{Cyclotron line energy measurements of Cepheus X-4 }
\tablehead{\colhead{Observation} &\colhead{Telescope} &\colhead{MJD} &\colhead{Source flux (3--78 keV)} &\colhead{} &\colhead{Cyclotron Line} &\colhead{}  &\colhead{Ref$^{(c)}$}\\
& & & (erg cm$^{-2}$ s$^{-1}$) &  Energy (keV) & Width (keV) & Depth ($\tau$) & } 
\startdata
1988 April & GINGA       & 47263.5   &  $7.23 \times 10^{-9}$$^{(b)}$  & $30.2^{+0.5}_{-0.5}$  & $16.1^{+1.8}_{-1.8}$  & $2.88^{+0.11}_{-0.11}$ & [1] \\\\
2002 July  & RXTE        & 52450.7   &  $1.08 \times 10^{-9}$$^{(b)}$  & $30.7^{+1.8}_{-1.9}$  & $3.6^{+2.9}_{-1.5}$  & $0.7^{+0.3}_{-0.2}$     & [4] \\\\
2014 June  & NuSTAR      & 56827.38  &  $2.69 \times 10^{-9}$  & $30.6^{+0.2}_{-0.3}$  & $3.9^{+0.2}_{-0.2}$  & $6.6^{+0.9}_{-0.8}$ $^{(a)}$     & [5] \\
2014 July  &             & 56839.87  &  $7.10 \times 10^{-10}$  & $29.0^{+0.3}_{-0.2}$  & $3.8^{+0.2}_{-0.2}$  & $7.0^{+0.8}_{-0.6}$ $^{(a)}$    & \\\\
2018 July  & AstroSat    & 58301.618 &  $1.62 \times 10^{-9}$  & $30.48^{+0.33}_{-0.34}$  & $3.24^{+1.14}_{-0.93}$  & $1.03^{+0.14}_{-0.11}$    & [6] \\
2018 July  &             & 58307.402 &  $8.08 \times 10^{-10}$  & $30.68^{+0.45}_{-0.44}$  & $5.89^{+1.38}_{-1.26}$  & $1.42^{+0.18}_{-0.18}$   &  \\\\
\hline
\enddata
\tablecomments{Errors quoted with 90\% confidence range for each parameter.\\ 
$^{(a)}$ expressed in terms of  absorption depth (d) related to optical depth $\tau = d/\sigma~(\sqrt{2\pi})$.\\
$^{(b)}$ Corrected flux with some approximation in 3-78 keV energy band.\\
$^{(c)}$References [1] Mihara et al. (1991) [4] McBride et al.(2007) [5] Vybornov et al.(2017) [6] Present Work}
\end{deluxetable}
%
%
\begin{figure}[h]
\centerline{\includegraphics[scale=0.36,angle=-90]{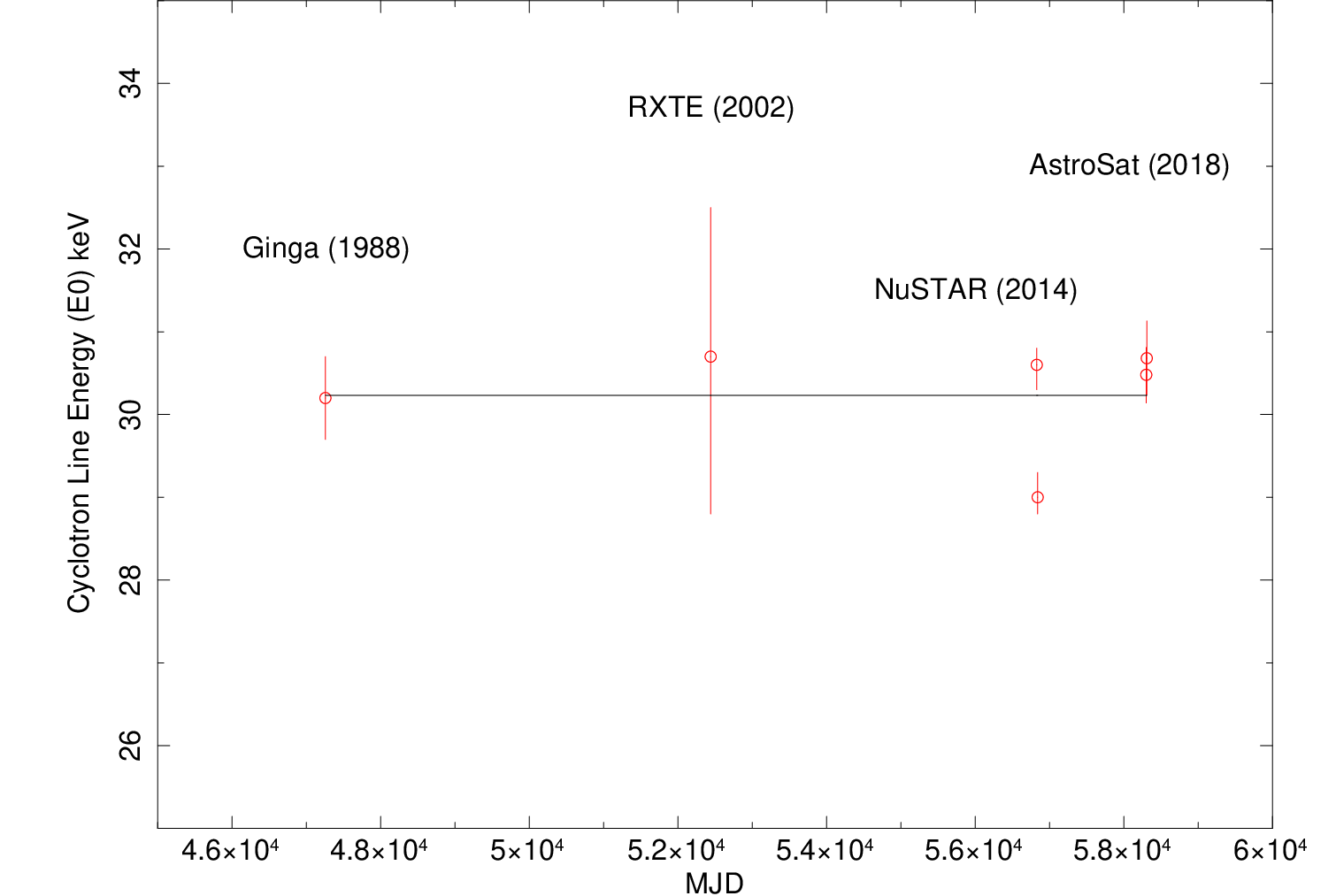}
\quad\includegraphics[scale=0.36,angle=-90]{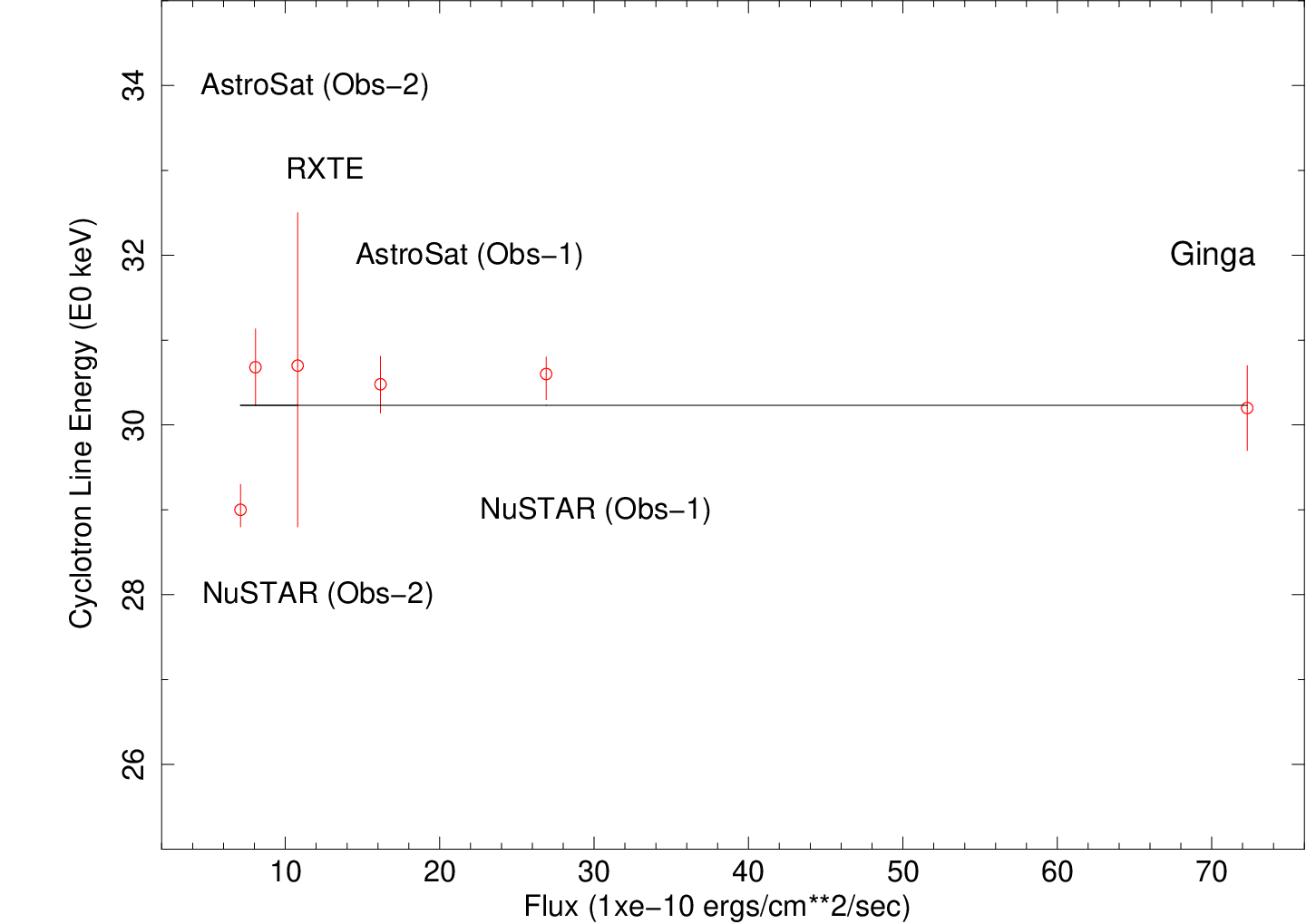}}
\caption{Cyclotron line energy of Cepheus X-4 measured  during different outbursts observed by 
different observatories over the period, Ginga(1988), RXTE(2002), NuSTAR(2014) 
and AstroSat(2018) respectively. (left-panel).  The cyclotron line energy plotted against the  
corresponding source flux as in Table~3 is also  shown  (right-panel). 
The horizontal line denotes the average value which corresponds
to 30.23 $\pm0.22$ keV in both the panels.  }
\label{lhist}
\end{figure}
%
%
The  centroid energy of the cyclotron line was plotted against the observed source flux 
as shown in Figure~\ref{lhist} (right-panel). It is  evident from the plot that  there are no specific 
trend observed  with respect to the source flux or luminosity.  This is in contrast to some of the  other wind-fed  
X-ray pulsars like, GX 304--1 \citep{furst15} and  Vela-X-1 \citep{laparola16} 
where they showed increase in line energy with increasing source luminosity.  For other pulsars such as 
V0332+53 \citep{makishima90}, 4U 0115+63 \citep{mihara95} and SMC X-2 \citep{klus14, lutovinov17} 
they showed decrease in line energy with increase in the source luminosity. In contrast to these
variations, cyclotron line energy of Cen X-3 showed a long duration stability with an average at 
$31.6\pm0.2$ keV over the period of 14 years \citep{ji19}.  Two observations by  NuSTAR  
of Cepheus X-4  during  2014 showed increasing trend with source luminosity \citep{furst15, vybornov17}. 
However,  when it is compared from other available measurements of Cepheus X-4, it did not show any 
such dependence on source luminosity as can be seen from Figure~\ref{lhist} (right-panel).  We also plotted  
cyclotron line energy   with respect to the time of observation, expressed in MJD as shown in 
Figure~\ref{lhist} (left-panel).  The plot does not show any specific trend with time or its luminosity. 
However some  significant variation in the  cyclotron line energy was observed with respect 
to its average value $30.23\pm0.22$ keV during 2014  outburst when source flux was found below 
about $8\times 10^{-10}$ erg cm$^{-2}$ s$^{-1}$ as can be seen in Figure~\ref{lhist} (right-panel).   
The average value of the cyclotron energy measurement is shown by the horizontal line. 
With respect to average value of cyclotron line energy, the AstroSat observation showed 
consistency  during 2018 outburst, in contrast to NuSTAR second observation of 2014, where it showed decrease. 
Relatively large errors associated with  the measurement by RXTE prohibit us to identify long term trend 
in between Ginga and NuSTAR observations and extent of variation more accurately.  
The flux measurement of RXTE in 3.5--10 keV energy band of 
$3.6 \times 10^{-10}$ erg cm$^{-2}$ s$^{-1}$ \citep{mcbride07} and 
similarly for Ginga measurements  of flux  $5.85 \times 10^{-9}$ erg cm$^{-2}$ s$^{-1}$ 
in 2--20 keV are under-estimate of actual flux in equivalent energy band of NuSTAR and AstroSat. 
We, therefore, scaled these fluxes by a  respective factor established by determining the ratios of fluxes from
equivalent energy band using the AstroSat spectrum. This corrections approximate the  corresponding fluxes 
from Ginga and RXTE in  3--78 keV energy band for comparison as given in Table~3.  The errors on these  estimates 
could only affect  a marginal shift of this point along the X-axis  corresponding to the  actual flux.  
The large error associated with the RXTE measurement of cyclotron line energy, do not indicate any 
significant trend even when flux is known accurately, similarly, Ginga measurements is close to the average 
value and hence its accurate flux estimate would not lead to  any possible trend.

The change in cyclotron line energy with luminosity for X-ray pulsars are  due to the effect of change in 
the behavior of the  line forming regions in the accretion column. It was shown by \citet{becker12} that when accreting
X-ray pulsars attain a critical luminosity $L_{crit}$ (Equation-12), then the radiation pressure in the accretion column 
decelerates  in-falling matter, thus affecting height of the cyclotron line forming region. 
The physical model  of \cite{becker12} also demonstrated that when  the source luminosity is higher than $L_\mathrm{crit}$ 
the line forming region in the accretion column shift to higher heights from the neutron star surface 
due to dominant radiation pressure of its supercritical luminous X-ray pulsar.  In contrast to this, when the source 
luminosity is below the $L_{crit}$, a sub-critical luminosity, the  cyclotron line energy increases due to 
the consequence of dominant pressure of  in-falling matter where the characteristic emission height decrease 
towards the surface of the neutron star. The sign of this dependence is thus opposite in the 
supercritical and sub-critical cases, hence creating the observed bimodal behavior for many X-ray pulsars.
\begin{equation}
  L_{crit} \approx 1.49 \times 10^{37} \left(\frac{\Lambda}{0.1} \right)^{-7/5} \left(\omega\right)^{-28/15} 
{\left(\frac{R}{10~km} \right)}^{1/10} \left(\frac{M}{1.4 M_\odot} \right)^{29/30} \left(\frac{B}{10^{+12}~G~} \right)^{16/15} 
	\mbox{~~erg~s}^{-1}.
	\label{ecrit}
\end{equation} 
The  constant 
$\Lambda$ in Equation-12  accounts  for certain  uncertainties such as due to spin-averaging of the magnetospheric radius,  
effects of plasma shielding  and other  magnetospheric effects. Whereas the  constant $\omega$ depends on the shape of 
the spectrum \citep{becker12}.  

The studies  of luminosity dependence for Cepheus X-4 from 2014 outburst data \citep{furst15}, suggested 
positive correlation with its luminosity.  \citet{vybornov17} conducted pulse-to-pulse analysis of the same 
NuSTAR data and also demonstrated positive correlation with source luminosity during the short time scale. 
We  compiled  all the   measurements of cyclotron line energy from  phase averaged spectra 
in Table~3 and presented in Figure~\ref{lhist} (right-panel) for  long term comparison. From these 
results, one can infer that 
there are no specific long term variation trend with respect to  source luminosity or time except  
with  an indication  that below a certain source flux $\sim 8 \times 10^{-10}$ erg cm$^{-2}$ s$^{-1}$ 
the  cyclotron line energy showed some  variability. For higher than this flux, cyclotron line energy 
almost remained stable close to its average value at $30.23\pm0.22$ keV.    
Now, for Cepheus X-4 the critical source luminosity estimated using Equation-\ref{ecrit}
is ${{\sim} 4 \times  10^{37}}$ erg s$^{-1}$. The source luminosity during two 
AstroSat observations were  estimated to be $2.04 \times 10^{37}$~erg~s$^{-1}$  
and  $1.02 \times 10^{37}$~erg~s$^{-1}$ respectively, for its  distance of 10.2 kpc 
as reported from Gaia observations \citep{malac20}. Both of these values are below the estimated value of 
$L_\mathrm{crit}$, which may have some uncertainty depending on the mass($M$) and radius($R$) of the neutron star and constants
$\Lambda$=0.1 and $\omega$=1 as considered here. The AstroSat
measurements should have shown  a positive correlation in cyclotron line energy with source luminosity with increase 
by a factor of $\sim 2$ 
as per the physical model of \citet{becker12}. But within its error limits, it showed almost similar and stable 
at its long term average value.  The Ginga observation is higher than the critical luminosity, where as the first observation 
of  NuSTAR is close to its critical luminosity. However, within the measurement errors cyclotron line energy shows
almost stable and  close to its long term average of $30.23\pm0.22$ keV rather than a negative correlation with source luminosity
as predicted by the model.  For more clarity, we further investigate physical parameters of accretion column 
and its implications.

The maximum height of the line forming region can be estimated  by the  equations described by \citet{becker07}, 
for a  super-critical case,  and  is given by
\begin{equation}
  h_{s} = 2 r_{0} \lesssim 1.3 \times 10^{5} 
\left(\frac{B}{10^{12}~G~} \right)^{-2/7} 
{{R}_6}^{9/14} \left(\frac{M}{M_\odot} \right)^{1/14} 
	\left(\frac{L_{37}}{1.3} \right)^{1/7}~~\hbox{cm},
\label{ehs}
\end{equation} 
where ${h_{s} = 2 r_{0}}$ where $r_{0}$  is polar cap or accretion column radius of the pulsar. 
The height of the bottom of the line-forming region is treated as the height of its thermal mound \citep{becker07}.    
The  height-size depends on source luminosity ($L$), the magnetic field ($B$), mass ($M$) and radius ($R$) of the neutron star.
Using the parameter values from Table~3 for Ginga and NuSTAR observation-1, the source luminosity
were  found to be $ {9.1 \times  10^{37}}$ erg s$^{-1}$ and ${3.4 \times  10^{37}}$ erg s$^{-1}$ respectively 
for source distance of 10.2 kpc.  Hence these observations can be treated as comparable to  super-critical case within
some approximation and their corresponding maximum height of line forming region can be estimated from above Equation-13 
as 1.2 km and 1.1 km respectively. 

The  height of the line-forming region ${h_{c}}$ in the accretion column for sub-critical luminous accreting X-ray pulsars, 
can be estimated using Equation-14.  For sub-critical luminous sources, the line forming
height reduces with increase in source luminosity where as for super-critical luminous sources  it increases with
increase in source luminosity as can be seen from given equations \citep{becker12}
\begin{equation}
  h_{c} =  1.48 \times 10^{5} \left(\frac{\Lambda}{0.1} \right)^{-1} \left(\frac{\tau}{20} \right) 
\left(\frac{M}{1.4 M_\odot} \right)^{19/14} {{R}_6}^{1/14} 
\left(\frac{B}{10^{12}~G~} \right)^{-4/7} {L}_{37}^{-5/7}~~\hbox{cm}.
\end{equation} 
For most of the accreting pulsars,  the  values of the parameters taken are ${\Lambda = 0.1}$ and 
Thomson optical depth ${\tau = 20}$. 
Where as  for other parameters of equations, values are  
obtained from the two AstroSat observations, we thus estimated the radius of accretion column  and
height of the line forming region above the thermal mound. Using  measured values of  
${B = 3.42 \times 10^{12}}$ G and ${L = 2.04 \times 10^{37}}$ erg s$^{-1}$  for observation-1 and    
${B = 3.40 \times 10^{12}}$ G and ${L = 1.02 \times 10^{37}}$ erg s$^{-1}$  for observation-2, we
estimate  size of the line forming regions  in accretion column for the sub-critical case,  during the two AstroSat observations as 
${h_{c} = 440}$ m and ${h_{c} = 725}$  m, respectively for the two observations. 
This indicate a difference of height of $\sim 285$ m between two observations.

The altitude, $H$, of the radiation-dominated shock can be estimated  in these cases using the expression 
of \citet{becker12}
\begin{equation}
  H_\mathrm{shock} =  1.14 \times 10^{5} \left(\frac{M}{1.4 M_\odot} \right)^{-1} {{R}_6} 
  {L}_{37}~\hbox{cm.}
\label{hshock}
\end{equation} 
 The respective  shock heights during AstroSat observations are 2.3~km and   
1.2~km, corresponding to two different luminosities.  From these 
estimates, it suggests that during the first observation of the AstroSat at higher source luminosity, 
the line-forming region height of $\sim 400$ m and its shock height is located at 2.3 km altitude in the accretion column, 
which is about 6 times away from the neutron star surface. Therefore, there are less chance of any possible 
impact of shock region on the line forming region, that may cause change in its size and location. Hence,
the cyclotron line energy remains almost stable similar to the values estimated at higher luminosities  by NuSTAR
and Ginga as can be seen in the  Figure~\ref{lhist} (right-panel). For the second AstroSat observation at lower luminosity, size
of the line forming region  is relatively larger $\sim 700$ m and shock height is about 1.2 km, which is 
relatively in closer proximity.  As there could be  uncertainties involved with  equation parameters, hence    
there is  possibility of some impact of shock region with the line forming 
region at this lower luminosity, which possibly may cause some variations in the cyclotron line energy depending 
on whether it may affect its size or its location or both.  However no such  effect is seen  for  the  
observation-2, at lower source luminosity of Cepheus X-4 by  AstroSat during 2018 outburst.  Now, for the NuSTAR 
observation-2, for which the  strength of magnetic field and source luminosity was measured as 
${B = 3.25 \times 10^{12}}$ G and ${L = 0.89 \times 10^{37}}$ erg s$^{-1}$ respectively.  The corresponding height of the 
line forming region was determined from Equation-14 as ${h_{c} = 818}$ m  and shock height using Equation-15 of
about 1.0~km are found to be in very close consistency considering parameters uncertainties involved.  
This may therefore, expected to  cause some variation  
in cyclotron line energy due to possible  interaction of shock region with the cyclotron line forming region.
The physical model of \citet{becker12} for sub-critical luminosity predict that cyclotron line energy
should decrease for NuSTAR observation-2.   
This is, therefore, possibly a very likely cause of sudden deviation or decrease  of cyclotron line  energy 
at ${E_{0}=29.0^{+0.2}_{-0.3}}$ keV during NuSTAR observation-2 from its  average energy of ${30.23 \pm 0.22}$ keV.   
This is when compared with super-critical cases the altitude of shock heights for Ginga and NuSTAR observation-1
are found be  about 10.3 km and 3.9  km as determined from Equation-15.  These are much beyond the corresponding  maximum heights
of line forming regions  of 1.2 km and 1.1 km as estimated above for a super-critical case.  This is therefore the
most possible reason that the cyclotron line energy  of Cepheus X-4 remains stable for wide range of variation 
of its luminosity.      
\citet{nishi14}  studied cyclotron line variation with source luminosity  based on changes in polar cap
dimensions, direction of propagation and changes of shock height with source luminosity. He successfully
explained observed changes, for  some of the wind-fed  X-ray pulsars using his model.
Interestingly, it was demonstrated  that  for a luminosity range  of ${\geq 1.0 \times 10^{37}}$ erg s$^{-1}$
line forming region remains almost same in terms of its dimensions and location, which do not change considerably
with luminosity this was particularly demonstrated for  the  X-ray pulsar X0115+63. This result is similar to 
the observations of Cepheus X-4, where it showed  stability of cyclotron line energy  for higher source luminosity.

\subsection{Pulse Phase-resolved Spectra}

Pulse-phase resolved spectroscopy for Cepheus X-4 was  first performed  with Ginga 
observations during  1988 outburst, for studies of the physical condition of line forming regions.  
An absorbed power-law with cyclotron scattering cut-off model was applied to the spectral data 
\citep{mihara91}.  It was subsequently followed up  during 2014 outburst  observed by  
NuSTAR, where an  absorbed Fermi-Dirac cutoff model combined with a black-body, 
an Fe-emission line and a cyclotron absorption line was employed for studies \citep{bhargava19}.
AstroSat observation of 2018 outburst of Cepheus X-4 was used for the most  recent studies of
phase resolved spectroscopy. A combined model defined by an absorbed compTT-model with 
a Gaussian for Fe-emission line along with a cyclotron absorption line defined by cyclabs-model was 
employed for 5-spectra  derived from 5-different  pulse phases for studies.
Results obtained  from the two AstroSat observations are presented  along with its spin-pulse overlay 
in Figure~\ref{phs2}, with the source luminosity differing by a factor $\sim 2$.  The AstroSat results
are also compared with earlier results in the following discussion below.  

The variation in spectral continuum and cyclotron line parameters were estimated over the pulse phase.
This was done by  measuring difference in the parameter range (maximum -- minimum) 
and expressed in percentage with respect to its phase-averaged value shown in Table~1. 
The 1-$\sigma$ error in percentage was determined by considering  respective 1-$\sigma$ error in
3 parameters considered.  The parameter variation is significant,
if the variation is found to be more than 3-$\sigma$ confidence limit.    
Thus, variation in cyclotron line energy for two AstroSat observations were found as 
$8.0\pm2.5\%$, $11.2\pm2.0$\%, the line-width as $105.0\pm30.4$\%, $85.0\pm19.0$\% 
and its optical-depth as $86.5\pm40.8$\%, $59.4\pm54.8$\% respectively. The variation in spectral
parameters such as soft photon Wien temperature $T_{0}$  were found as 
$22.5\pm19.0$\%, $45.8\pm20.5$\% and plasma temperature $kT$ as $48.2\pm26.0$\%, 
$58.6\pm28.3$\% and for its associated optical-depth as $48.7\pm5.1$\%,  $60.9\pm5.9$\% respectively,   
for two AstroSat observations. This shows that only component of 
the spectral continuum defined by compTT-model, plasma optical-depth 
showed significant variation over the pulse phase for both AstroSat observations.  While cyclotron line
energy showed  considerable and  relatively higher variation for observation-2 than observation-1
for AstroSat, it is not statistically significant though. 
The cyclotron line-width showed considerable variation for two observations,
whereas,  the variation in line-depth was not statistically  significant  for both of the
observations within error limits.            
This when compared to Ginga results, we find variation in line-energy of $7.2\pm1.4$\%,
and in line-width as $72\pm6$\% whereas in its line optical-depth as $54.6\pm5.3$\%.  
The source luminosity was relatively much higher for Ginga \citep{mihara91} as can be seen in the Table~3.   
It was noticed that the  variation in line-width of the cyclotron line  was though higher for 
two AstroSat observations, but the measurements errors were also relatively higher than that of 
the Ginga observation. The variation of the optical-depth of the cyclotron line  
during two AstroSat observations are though comparable to Ginga, but due to larger errors
associated with AstroSat, we are not able to detect any significant change during these observations. 
The results from two NuSTAR observations in 2014, showed  variations in cyclotron line energy 
of $15.4\pm4.2$\%, $8.1\pm2.2$\% when measurement for observation-2 with very large error was 
dis-regarded and next minimum was considered.  
The cyclotron line-width variation of  $83.3\pm2.4$\%, $94.3\pm13.6$\% and optical-depth 
of $211.0\pm24.0$\%,  $176.0\pm27.1$\%, were  observed respectively for two observations of NuSTAR, 
at  different source flux (or luminosity) as given in Table~3. 
As  the  pulse-phase was divided into 9 different phases for NuSTAR data, therefore, 
these variations are higher compared to AstroSat results, where pulse-phase was divided into 
5 longer phase intervals and resultant variations were relatively lower  due 
to effect of phase averaging. There could also be a marginal effect  due to selection of different
continuum model particularly for Ginga results.  Therefore, direct comparison of these variations may not be appropriate. 
However, one can compare observed  trend of these variations of cyclotron line parameters with
pulse-phase.  A  similar  trend of variation of the  centroid energy of cyclotron line was noticed,  
for  AstroSat 2018 data as that of NuSTAR 2014  observations. It remains almost  stable  
within error limits except  within the rising  and the falling edge of the primary-pulse of AstroSat, 
where these variations  are consistent with respect to  NuSTAR observation-2 at lower 
source luminosity. 
The cyclotron line-width variations trend almost followed
pulse intensity, for both the AstroSat observations. 
This variation  from NuSTAR results  was found to be very similar to observation-1 than that of observation-2.
AstroSat observation of cyclotron line optical-depth variation  for both observations 
are found almost stable  with  the spin-pulse 
intensity variation except in the pulse minima where it showed increasing trend.
The optical-depths of cyclotron line  for the  AstroSat observation-2 showed  marginally higher  values 
for all the phases  than the observation-1. 
The NuSTAR  measurements of optical depth  appeared to show variation with spin-phase for observation-1 within its large error
limits, however for  observation-2, during the rising phase of the secondary pulse,  it remains  almost same at relatively  
lower value. 
Thus, AstroSat phase resolved spectroscopy  although showed overall a  similar trend of  variations except 
with distinct differences corresponding to certain phases where extent of relative variations are different.

Cepheus X-4 showed overall variation of 5--10\% in cyclotron line-energy  
over the pulse-phase. In contrast to this, other sources such as Cen X-3 \citep{burderi00, suchy08}, 
Vela X-1 \citep{barbara03, kreykenbohm02}, 4U 0115+63 \citep{heindl04},
GX 301--2 \citep{kreykenbohm04, suchy12},  Her X-1  \citep{voges82, soong90, klochkov08} and 
GRO J2058+42 \citep{mukerjee20}, showed  variation in their fundamental cyclotron line energy of about 
10--30\% over the pulse-phase. 
The variation in cyclotron features with pulse-phase may occur due to overall effect of 
superposition of a number of lines formed at different heights of line-forming region in its 
accretion column, affected by a large field gradient of neutron star magnetic field. 
Therefore, variation of the centroid energy and shape of the cyclotron absorption feature may
be possible depending on orientation of the overall system and visibility of accretion column 
with respect to line of sight \citep{nishi15}.  
The variation in spectral continuum as well as in the cyclotron lines, may occur  
due to influence of the velocity of bulk motion of in-falling plasma in the  
line-forming regions as explained in the recent work by \citet{nishi19}.  As line forming  regions are located  near the 
walls of the accretion column of its cylindrical geometry and spectral continuum are 
formed above and around the accumulated mound close to surface of the neutron star.  
Therefore,  based on the relative location of these two regions with 
respect to an observer line of sight, variation  in the spectral 
continuum as well as variation in cyclotron-line  parameters could be observed during certain  pulse-phase 
depending on optimum conditions of the  overall system such as velocity of the bulk motion of the in-falling 
matter, gravitational effects on the emitted radiation and  local variation of its magnetic field \citep{nishi19}.  
The neutron star surface area where magnetic field is more intense,  may produces a hotter region resulting 
in a relatively hotter plasma accumulation.  These may cause a significant variation in the spectrum 
due to the interaction of accelerated high energy particles with soft photons which give rise to the process of 
Comptonization, resulting change in the spectral shape with pulse phase.
It is therefore inferred  that the observed  phase dependent changes in the spectral features 
as well as in the  parameters of the cyclotron line for Cepheus X-4 during AstroSat observation, 
could likely be  due to such  changes such as in the geometry, physical conditions in the accretion column,  
local magnetic field strength of the line producing region  and  the system orientation with respect to 
the observer line of sight during its spin phase, at a particular source luminosity.

The cyclotron line  energy of Cepheus X-4 within 90\% confidence limit showed variation  with  
its pulse phase in a range of (28.5--32.9) keV.  This indicate  a variation in 
the magnetic field strength of (3.2--3.7)$\times 10^{12}$ G over the pulse phase of the pulsar
including the gravitational effect of its neutron star.
The magnetic field strength derived from its spectrum for Cepheus X-4 is 
comparable to the magnitude of other X-ray pulsars, such as  X-Persei ($28.6^{+1.5}_{-1.7}$ keV) \citep{coburn01},   
RX J0440.9+4431 ($31.9^{+1.3}_{-1.3}$ keV) \citep{tsygan12}, 
RX J0520.5-6932 ($31.3^{+0.8}_{-0.7}$ keV) \citep{tendul14} 
and Cen X-3 ($31.6^{+0.2}_{-0.2}$ keV) \citep{ji19}. 
where energy values of their respective fundamental 
cyclotron absorption features are shown within parenthesis.

\section{Conclusion}

Timing and spectral properties of Cepheus X-4 were studied during declining phase of 2018 outburst,
using  AstroSat observations at two different source luminosities.  
The pulsar was observed  in spin-down phase  during both the observations.  Spin-period 
and spin-down rate  of the pulsar were determined as $65.35080\pm0.00014$ s, 
$(-2.10\pm0.8)\times10^{-12}$ Hz s$^{-1}$ at  MJD 58301.61850 and   
$65.35290\pm0.00017$ s,  $(-1.6\pm0.8)\times10^{-12}$ Hz s$^{-1}$ at MJD 58307.40211 for two observations.  
Pulse-profiles derived  at different energy bands covering 0.5--60 keV, showed pronounced and  
multiple structures at lower energies,  and  showed a variation in shape and pulse-fraction 
with energy.  The RMS pulse fraction showed variations  with  energy and source 
luminosity. After certain  higher luminosity of the source, pulse fraction  showed  
overall decrease.  Pulse-profiles derived from recent AstroSat observations were found  
to be similar in shape to those reported earlier from RXTE  1997 and NuSTAR 2014 observations.
The pulsar between the outburst showed continuous spinning down  at an average rate  
of $(-2.455\pm0.004)\times10^{-14}$ Hz s$^{-1}$ over last 30 years, due to its  entry into 
subsonic propeller regime.  
Spectral measurements  between 0.7--55 keV showed difference in its continuum for 
two different source luminosity for phase-averaged spectrum. 
The prominent cyclotron resonance scattering features with a peak absorption energy of
$30.48^{+0.33}_{-0.34}$ keV and $30.68^{+0.45}_{-0.44}$ keV for FD-cutoff model and
$30.46^{+0.32}_{-0.28}$ keV and $30.30^{+0.40}_{-0.37}$ keV for compTT-model were detected.
The  cyclotron line energy showed some deviations below a specific source luminosity, otherwise
remained stable at an average value of 30.23 $\pm$ 0.22 keV established from our studies.   
The pulse-phase dependent variations were observed in some of the parameters of  spectral continuum  and 
cyclotron line. These variations were more for the lower luminosity of the source.
The detection of cyclotron line  lead to determination of strength of its strong 
magnetic field of the neutron star.  Therefore, using AstroSat observation, 
we could establish the variation of strength  of magnetic field over the pulse-phase 
of the pulsar as  (3.2--3.7)$\times 10^{12}$ G.
Further, follow up observation of the source during its  rare outbursts  would
offer us scope for study  spectral variability with source 
luminosity, variation of magnetic field strength with time and to probe nature 
of its accretion column  and geometry where cyclotron lines are produced. 

\acknowledgments

We gratefully acknowledge  vital contributions and crucial support of the Indian Space 
Research Organization (ISRO) for the success of the AstroSat mission.  The generous  
support of ISRO for instrument building, tests and qualifications, software developments and 
mission operations are always gratefully acknowledged.  We thankfully acknowledge the 
support of respective Payload Operation Centers (POCs) of the Large Area Proportional 
Counter (LAXPC) as well as the Soft X-ray Telescope (SXT) at TIFR Mumbai, for the release 
of verified data, calibration data products and pipeline processing tools. 
We gratefully acknowledge the support of the Department of Atomic Energy, Government of India, 
under project~no.~12-R\&D-TFR-5.02-0200. We  acknowledge generous support through 
the High Energy Astrophysics Science Archive Research Center On-line Services, for the 
data obtained from Swift-BAT provided by the NASA/Goddard Space Flight Center for this
research work.  We gratefully acknowledge  generous support of  NASA's HEASARC  for offering 
all the useful software  and tools  for analysis of  Astronomical data. Last but never 
the least, we express our sincere thanks to the referee for very valuable and constructive
comments to improve the paper.  

\facility{Astrosat}

\end{document}